# Technical Report # KU-EC-15-1:
# Analysis of Cortical Morphometric Variability Using Labeled Cortical Distance Maps

E. Ceyhan[1*], T. Nishino[2], K.N. Botteron[2,3], M.I. Miller[4,5,6], J.T. Ratnanather[4,5,6]

[1]*Dept. of Mathematics, Koç University, 34450, Sarıyer, Istanbul, Turkey.*
[2]*Dept. of Psychiatry, Washington University School of Medicine, St. Louis, MO 63110.*
[3]*Dept. of Radiology, Washington University School of Medicine, St. Louis, MO 63110.*
[4]*Center for Imaging Science, The Johns Hopkins University, Baltimore, MD 2121*
[5]*Institute for Computational Medicine, The Johns Hopkins University, Baltimore, MD 21218.*
[6]*Dept. of Biomedical Engineering, The Johns Hopkins University, Baltimore, MD 21218.*

*corresponding author:
Elvan Ceyhan,
Department of Mathematics, Koç University,
Rumelifeneri Yolu, 34450 Sarıyer, Istanbul, Turkey
e-mail: elceyhan@ku.edu.tr
phone: +90 (212) 338-1845
fax: +90 (212) 338-1559

**short title:** Morphometric Variability with Labeled Cortical Distance Maps
**keywords:** Brown-Forsythe test, censoring, computational anatomy, homogeneity of variance, pooled distances, simultaneous inference

## ABSTRACT
Morphometric (i.e., shape and size) differences in the anatomy of cortical structures are associated with neuro-developmental and neuropsychiatric disorders. Such differences can be quantized and detected by a powerful tool called Labeled Cortical Distance Map (LCDM). The LCDM method provides distances of labeled gray matter (GM) voxels from the GM/white matter (WM) surface for specific cortical structures (or tissues). Here we describe a method to analyze morphometric variability in the particular tissue using LCDM distances. To extract more of the information provided by LCDM distances, we perform pooling and censoring of LCDM distances. In particular, we employ Brown-Forsythe (BF) test of homogeneity of variance (HOV) on the LCDM distances. HOV analysis of pooled distances provides an overall analysis of morphometric variability of the LCDMs due to the disease in question, while the HOV analysis of censored distances suggests the location(s) of significant variation in these differences (i.e., at which distance from the GM/WM surface the morphometric variability starts to be significant). We also check for the influence of assumption violations on the HOV analysis of LCDM distances. In particular, we demonstrate that BF HOV test is robust to assumption violations such as the non-normality and within sample dependence of the residuals from the median for pooled and censored distances and are robust to data aggregation which occurs in analysis of censored distances. We recommend HOV analysis as a complementary tool to the analysis of distribution/location differences. We also apply the methodology on simulated normal and exponential data sets and assess the performance of the methods when more of the underlying assumptions are satisfied. We illustrate the methodology on a real data example, namely, LCDM distances of GM voxels in ventral medial prefrontal cortices (VMPFCs) to see the effects of depression or being of high risk to depression on the morphometry of VMPFCs. The methodology used here is also valid for morphometric analysis of other cortical structures.

## 1  Introduction
Quantification of morphometric properties of cortical structures is a major component of Computational Anatomy (CA). Recently, the laminar structure of the neo-cortex has received considerable attention thanks to advances in high resolution magnetic resonance imaging (MRI) technology and the development of CA methods (see, e.g., [1-6]). Specifically, Labeled Cortical Distance Mapping (LCDM) has been used for structural comparisons of cortical thickness in the cingulate cortex in studies of Alzheimer's disease [7] and schizophrenia [40] in comparison to control subjects. The LCDM algorithm provides distances of labeled gray matter (GM) voxels from the GM/white matter (WM) surface for cortical structure of interest. Depending on the resolution of the GM voxels, the associated data set can be very large for each subject. Previously, the morphometric differences between the diagnostic groups were discovered by the analysis of LCDM distances, more specifically, by comparing the means and distributions of the LCDM distances [8, 9]. However, in this article, cortical morphometric variability (possibly due to a disease or impairment) as measured by the LCDM distances is studied.

Cortical thinning has been observed in other regions in a variety of neuro-developmental and neuro-degenerative disorders (see above references for examples). In particular, functional imaging studies implicate the ventral medial prefrontal cortex (VMPFC) in major depressive disorders (MDD) [10, 11] which have been correlated with shape changes observed in structural imaging studies [12, 13]. Some specific regions of the prefrontal cortex play an important role in modulating emotions and mood. Structural imaging studies in MDD have largely focused on adult onset with only few focused on early onset MDD which has been associated with structural deficits in the subgenual prefrontal cortex, a subregion of the VMPFC. Furthermore, the whole VMPFC has been examined in a twin study of early onset MDD [13].

Several studies of the VMPFC and related structures have been obtained from analysis of the cortex as a whole [13-17], whereas others have pursued more attempts at the localized analysis attempts to deal with the highly folded and variable GM cortex [18] and to address issues of signal inhomogeneity or artifacts which can cause processing issues in this region. Also in the localized approach, the laminar shape (i.e., curvature/folding and thickness) of the cortex can be quantified in great detail. Two aspects of the laminar shape are structural formation (like surface and form of the cortex or curvature of the tissue) and scale or size (like volume and surface area). Thus, *morphometry* refers to all aspects of laminar structure, where "shape" refers to the surface structure and "size" refers to the scale of the tissue in question.

LCDMs can be used in many ways. For example, the $90^{th}$ percentile distance of LCDMs can be used as a measure of cortical thickness [19]. Group comparisons can be performed on GM volumes (by the usual *t*-test) or on randomly selected subsamples from LCDM distances can be analyzed via Wilcoxon rank sum (WRS) test [7]. Various morphometric measures (i.e., volume, descriptive statistics based on LCDM distances such as median, mode, range, and variance) can be used for group comparisons; however, these variables were shown to have less power in discriminating the depressed subjects from healthy ones, possibly due to oversimplification of cortical characterization represented in LCDM distances [9]. To avoid this information loss, the LCDM distances can be pooled (i.e., merged) for each diagnostic group and various statistical tests (such as tests on differences in distribution (such as Kolmogorov-Smirnov (KS) test or WRS test) or location (i.e., tests on mean or median such as AVOVA *F*-tests and *t*-tests) can be performed by classical parametric or non-parametric tests [8]. Furthermore LCDM distances can be censored at various (consecutive) distance thresholds to determine the location (i.e., distance from the GM/WM surface) where significant differences in morphometry starts [20]. That is, at each step we only use the voxels with distances smaller than the threshold distance and we do not use the voxels with distances larger than the threshold distance, hence the term "censoring" for this process. Previously, the distributional and mean differences at the diagnostic group level are investigated using LCDM distances [8, 9] and also the location (i.e., the distance from the GM/WM surface) of such differences were found by using censored LCDM distances [20]. But, in this article, the morphometric variability of the region of interest (ROI) due to the disease in question is assessed. Due to the structure of LCDM distances, the location, distribution, and variance comparisons might provide complementary information. In particular, LCDM distances are positive, and with a skewed right probability density function (pdf). Hence, the distributional and location analysis are not totally unrelated to the morphometric variability. Here, we investigate the similarities and differences between variability (or variance) analysis when compared to distributional and location analysis. The effect of a disease or a disorder on the variability may be important in the sense that, if, e.g., the disease reduces variability, then the diagnostic inference on the morphometry of the ROI will be more precise in indicating the possibility of the disease. However, variability analysis should be performed in addition to or in conjunction with the distributional or location analysis to obtain more reliable information.

The LCDM approach has been applied in clinical neuroimaging studies of the cingulate in subjects with Alzheimer's Disease [21] and schizophrenia [22-24], the prefrontal cortex in subjects with schizophrenia [25], the parahippocampal gyrus in subjects with schizophrenia [26], the occipital cortex in visual attention [27, 28], Area 46 of the frontal cortex in fetal irradiated macaques [29] and ERC in normal aging controls and in subjects with mild cognitive impairment [30]. Finally, our observation of variable cortical thickness in the left PT in three groups of age-matched and gender-matched controls and patients with schizophrenia and bipolar disorder [31] is consistent with post-mortem analysis [32]. The approach has also been extended to deal with deeply buried sulci by modeling image intensity stochastically based on the normal distance where the model includes cortical thickness as one of the parameters [1]; others have similarly adapted LCDMs [33, 34]. The LCDM approach is similar to the voxel-based cortical thickness method (VBCT) [35] where each voxel in the GM has a thickness value associated with it, but our analysis of these voxel-based thickness values is different. In VBCT cortical thickness values are compared on a voxel-by-voxel basis as in SPM2 [36], while our analysis of LCDM distances allows us, for example, to first pool (i.e., merge) the distance values for each diagnostic group, and perform the comparisons on the overall distance (or thickness) level, rather than the voxel level for each individual. It has been shown that LCDMs are comparable to other methods for computing cortical thickness [37] and that LCDM profiles for whole brains are similar in shape [35, 38]. LCDM is essentially a ROI approach and therefore differs from global ones such as FreeSurfer [39] which averages point-to-point distances between outer and inner cortical surfaces.

HOV analysis of pooled distances provides an overall analysis of morphometric variability of the LCDMs at the group level due to the disease in question, while the HOV analysis of censored distances suggests the location(s) of significant variation in these differences (i.e., at which distance(s) from the GM/WM surface the morphometric variability starts to be significant on the average) at the group level. We also check for the influence of the assumption violations on the HOV analysis of LCDM distances. In particular, we demonstrate that BF HOV test is robust to assumption violations by the LCDM distances such as non-normality and within sample dependence of the residuals from the median. Furthermore, at each censoring step, the censored distances aggregate, which might confound the results of statistical testing. We investigate the influence of data aggregation with a Monte Carlo simulation analysis and show that such influence on the HOV test is only mild and negligible in practice. We also assess the performance of the methods on normal and exponential data sets to see the performance of the tests when more assumptions than those of LCDM distances are satisfied.

As an illustrative example, we perform LCDM analysis of GM tissue in VMPFCs in a study of early onset depression in twins. The methodology is applied to LCDMs generated for the VMPFC implicated in major depressive disorders (MDD) [10, 11, 13, 18, 40]. Furthermore, in analysis of censoring distances, we have the multiple testing problem when all censoring thresholds are considered together. We consider various corrections for simultaneous inference (i.e., $p$-value adjustment or correction for multiple testing) and compare these correction methods. We demonstrate that among these methods, Benjamini-Hochberg correction [41] has the best performance.

We describe the acquisition of LCDM distances for VMPFCs in Section 2.1, pooling and censoring of LCDM distances in Section 2.2., statistical methods we employ in Section 2.3, assess the influence of assumption of violations and data aggregation with an extensive Monte Carlo simulation study in Sections 3.1-3.3, assess the performance of the methods on normal and exponential data in Section 3.4, and discuss the simultaneous inference procedures for analysis of censored distances in Section 3.5. We present the HOV analysis of the example data set in Section 4, and discussion and conclusions in Section 5.

## 2. Methods

## 2.1 Data Acquisition

A cohort of 34 right-handed young female twin pairs between the ages of 15 and 24 years old were obtained from the Missouri Twin Registry and were used to study cortical changes in the VMPFC associated with MDD. Both monozygotic and dizygotic twin pairs were included, of which 14 pairs were controls (Ctrl) and 20 pairs had one twin affected with MDD, their cotwins are designated as the High Risk (HR) group. Three high resolution T1-weighted MPRAGE magnetic resonance scans of each subject in this population were acquired using a Siemens 1.5T Sonata scanner with 1 $mm^3$ isotropic resolution. Images were then averaged, corrected for intensity inhomogeneity and interpolated to $0.5 \times 0.5 \times 0.5$ $mm^3$ isotropic voxels. Following Ratnanather et al. (2001), a ROI comprising the prefrontal cortex stripped of the basal ganglia, eyes, sinus, cavity, and temporal lobe was defined manually and segmented into gray matter (GM), white matter (WM), and cerebrospinal fluid (CSF) by Bayesian segmentation using the expectation maximization algorithm [42]. A triangulated representation of the cortex at the GM/WM boundary was generated using isosurface algorithms [43].

LCDM is a method to characterize the cortical laminar shape over a specified cortical ROI and are generated as follows: first, the ROI subvolume is partitioned by a regular lattice of voxels of specific size $h$, denoted $V(h)$. Each voxel is a cube of size $h \times h \times h$ (in some unit, say, $mm^3$). Every voxel is labeled by tissue type as gray matter (GM), white matter (WM), and cerebrospinal fluid (CSF) [e.g. 2, 42]). For every voxel in the ROI subvolume, the (normal) distance from the center of the voxel to the closest point on GM/WM surface is computed. Let $S(\Delta)$ be the triangulated graph representing the smooth boundary at the GM/WM surface. The distance computation algorithm is specified as follows [2, 18, 19]:

**for all** $i$ **do**
   $s_{\text{closest}} \leftarrow$ **a point in** $S(\Delta)$ **such that**
   **for all** $s_j \in S(\Delta)$ **do**
      $d\left(s_{\text{closest}}, \text{centroid}(v_i)\right) \leq d\left(s_j, \text{centroid}(v_i)\right)$
   **end for**
   $D_i \leftarrow d\left(s_{\text{closest}}, \text{centroid}(v_i)\right)$
**end for**

where $d(\cdot,\cdot)$ stands for the usual Euclidean distance, $v_i$ is the $i^{\text{th}}$ voxel, $s_j$ is the $j^{\text{th}}$ point in $S(\Delta)$ and $D_i$ is the $i^{\text{th}}$ distance (i.e., distance for the $i^{\text{th}}$ voxel). That is, an LCDM distance is a set distance function $d: v_i \in V \to d\left(\text{centroid}(v_i), S(\Delta)\right)$, which is the distance between the centroid (or center of mass) of $v_i$ and the set $S(\Delta)$. More precisely,

$$D_i := d\left(\text{centroid}(v_i), S(\Delta)\right) = \min_{s \in S(\Delta)} \left\| \text{centroid}(v_i) - s \right\|_2. \tag{1}$$

where $C_M(\cdot)$ stands for center of mass (or centroid), and $\|\cdot\|_2$ is the usual $L_2 - \text{norm}$. A signed distance is used to indicate the location of each voxel with respect to the GM/WM surface; distances are positive for GM and CSF voxels, and negative for WM voxels. These distances constitute the LCDM distance data. We will only use LCDM distances for the GM of the ROI in our further analysis. Hence, LCDM distances refer to those of GM voxels for the rest of the article.

Let $D_{ijk}^L$ be the distance associated with $k^{th}$ voxel in GM of left VMPFC of subject $j$ in group $i$ for $j = 1, 2, \ldots, n_i$, $i = 1, 2, 3$ (group 1 is for MDD, group 2 for HR, and group 3 for Ctrl) and $k = 1, 2, \ldots, m_{ij}^L$ where $m_{ij}^L$ is the number of voxels for left VMPFC of subject $j$ in group $i$. Thus, $n_1 = 20$, $n_2 = 20$, and $n_3 = 28$. Right VMPFC distances are denoted similarly as $D_{ijk}^R$ with $m_{ij}^R$ being the number of voxels for right VMPFC of subject $j$ in group $i$. Based on prior anatomical knowledge (e.g., [44]), cortical thickness of the VMPFC is roughly 6 $mm$ so we only retain distances larger than -0.5 $mm$ so that mislabeled WM is excluded from the data with an upper limit of 5.5 $mm$. This causes discarding a small portion of the LCDM distance data set. In particular, in our VMPFC LCDM distances data, it turns out that only 0.16% of left distances and 0.14% of right distances are below -0.5 $mm$; on the other hand, only 0.22% of left distances and 0.07% of right distances are above 5.5 $mm$.

By construction, most of GM distances are positive, most of WM distances are negative, and all of CSF distances are positive. Mismatch of the signs for some GM and WM voxels close to the GM/WM boundary are due to the way the surface is constructed in relation to how the pixels are labeled, such that a surface is always intersecting pixels, i.e., partial volume. Hence some appropriately labeled GM and WM pixels may fall on a side of surface that they should not belong to; however, these mislabeled voxels constitute a small number of voxels and do not affect the overall analysis. Reliability of LCDMs is dependent on reliability of GM segmentation and reconstruction of GM/WM surface which has been validated for several cortical structures including VMPFC [18], cingulate gyrus ([45]; [46]) and planum temporale [47].

## 2.2 Pooling and Censoring of LCDM Distances by Group

We pool LCDM distances of subjects from the same diagnostic group or condition; that is, we pool the LCDM distances of all left MDD VMPFCs in one group, those of all left HR VMPFCs in another group, and those of all left Ctrl VMPFCs in another. Let $S_{ij}^L = \{D_{ijk}^L, k = 1, 2, \ldots, m_{ij}^L\}$ be the set of LCDM distances of subject $j$ from group $i$ for $j = 1, 2, \ldots, n_i$ and $i = 1, 2, 3$. Then we set

$$D_i^L = \bigcup_{j=1}^{n_i} S_{ij}^L = \{\mathcal{D}_{i\ell}^L, \ell = 1, 2, \ldots, n_i m_{ij}^L\} \qquad (2)$$

where $D_i^L$ is the set of left distances for group $i$ with $i = 1, 2, 3$, $\mathcal{D}_{i\ell}^L$ is the $\ell^{th}$ distance value for the merged (or pooled) distances from left VMPFCs of subjects in group $i$ for $\ell = 1, 2, \ldots, n_i m_{ij}^L$. We pool the right VMPFC LCDM distances in a similar fashion and replace $L$ with $R$ in the notation.

One of the underlying assumptions for pooling is that the distances from left VMPFCs of subjects with the same diagnostic condition have the same distribution (i.e., distances of left VMPFCs of subjects with MDD have the same distribution, say $F_1^L$, those of HR have the same distribution, $F_2^L$, and so do those of Ctrl group, with $F_3^L$); and the same holds for distances of right VMPFCs with distributions $F_i^R$ for $i = 1, 2, 3$. In other words, we assume that $D_{ijk}^L$ are identically distributed for all $j = 1, \ldots, n_i$ and $k = 1, \ldots, m_{ij}^L$. So, $D_{ijk}^L \sim F_i^L$ for all $j$, $k$, and $S_{ij}^L$ is a sample from the distribution $F_i^L$; likewise $D_{ijk}^R \sim F_i^R$ for all $j$, $k$. Hence the pooled distances are distributed as $\mathcal{D}_{i\ell}^L \sim F_i^L$ and $\mathcal{D}_{i\ell}^R \sim F_i^R$ for $i = 1, 2, 3$ and $\ell = 1, 2, \ldots, n_i m_{ij}^L$. We take this action under the presumption that the morphometry of VMPFCs of the healthy subjects are similar and those of subjects with the same disease are affected in the same manner, hence age and gender matched subjects with the same health condition (whether healthy or diseased) have VMPFCs similar in morphometry. We also denote all left pooled distances as

$$D^L = \bigcup_{i=1}^{3} D_i^L = \{D_{ijk}^L, i = 1, 2, 3, j = 1, 2, \ldots, n_i, k = 1, 2, \ldots, m_{ij}^L\}$$

and similarly denote all right pooled distances as $D^R$ replace $L$ with $R$ in the notation.

Next, we partition the range of LCDM distances into bins of size $\delta$, then we have $\lfloor d_{\max}/\delta \rfloor$ many bins where $\lfloor s \rfloor$ stands for the floor of (i.e., largest integer less than or equal to) $s$. To construct LCDM censored distances, we only retain distances less than or equal to a specified distance value denoted $\gamma_{\delta,k}$. In particular, at step $k$, we only consider the voxels whose LCDM distances are less than equal to $\gamma_{\delta,k} = k\delta$. Thus we only consider the layer of the cortex with thickness of roughly $k\delta$ from the GM/WM surface. These distances are called the *censored LCDM distances*, which, for left VMPFCs, are denoted as

$$C_d^L(k, \delta) := \{d \in D^L \cap [-0.5, k\delta]\} = \{d \in D^L : d \leq k\delta\}$$

and for group $i$ of left VMPFCs,

$$C_{d,i}^L(k, \delta) := \{d \in D_i^L : d \leq k\delta\}$$

for $i = 1, 2, 3$ (i.e., for groups MDD, HR, and Ctrl, respectively). Censored LCDM distances for right VMPFCs are denoted similarly as $C_d^R(k, \delta)$ and for group $i$ of right VMPFCs as $C_{d,i}^R(k, \delta)$.

Censored distances depend on the bin size, $\delta$ and resolution of the voxels $h$. We recommend the use of a bin size between $[h/10, h]$ (which corresponds to 0.01 to 0.5 *mm* for our data). Because, if they are too large, censored distances do not provide the desired resolution in the distances from the GM/WM surface, and if they are too small, they do not improve on the results of 0.01 *mm* but increase computation time. So the lower bound on the bin size is rather of practical choice. In what follows, we use $d_{max} = 5.5$ *mm* and $\delta = 0.01$ *mm.* Therefore, we have $k = 0, 1, 2, ..., 551$ and $\gamma_{\delta,k} = .00, .01, .02, ..., 5.50$ mm. To overcome the possible confounding effects of the mislabeled GM voxels close to the GM/WM surface, censored distances within [0.5, 5.5] *mm* are used for statistical inference, as these censored distances provide more reliable results. Moreover, the censored distance analysis is performed with the same tests as the pooled distance analysis performed in [20].

## 2.3 Statistical Tests

Previously, we have analyzed the pooled and LCDM distances for differences in location and distribution in [8] and [20], respectively. In this article, we employ a test of equality or homogeneity of the variances (HOV) of pooled and censored distances which is important in its own right, because variance differences in distances between groups might be indicative of differences between the morphometric variations/variability of VMPFCs (due to the disease). That is, variance of LCDM distances for, e.g., left VMPFC of a subject is a measure of shape or size variation in that particular VMPFC.

HOV tests are usually employed to check an important assumption in ANOVA and the *t*-test for mean comparisons which states that the variances in the different groups are equal (i.e., homogeneous). The two most common HOV tests are the *Levene's test* and the *Brown-Forsythe test* where the latter is a modification of Levene's test. HOV assumption is not a crucial one for ANOVA methods, so we perform HOV tests not for assumption checking for ANOVA but for a different purpose: to determine the effect of a certain disease on the morphometric variability of a brain tissue. In Levene's test ANOVA is performed on the absolute deviations (called residuals) of the values from the mean, and BF test does the same but on deviations from the median. The basic assumptions for these HOV tests are the same as the ANOVA assumptions, not on the original variable but on the residuals from the mean or median. That is, the residuals should enjoy within sample independence, between sample independence, normality (i.e., Gaussianity), and equality of the variances (of the residuals). It has been shown that BF test is more robust to the assumption violations [48], hence we prefer it over Levene's test in our further HOV analysis of LCDM distances. Therefore, we perform HOV by using Brown-Forsythe's (BF) HOV test (see, e.g., [49]). We apply a multi-group BF HOV test and if this is significant, then we perform (directional) pairwise HOV comparisons with Holm's correction [50]. In the literature, BF HOV (and Levene's HOV) tests are only used for two-sided alternatives, as they are based on ANOVA *F*-test on the residuals. However, in the two-sample case, ANOVA *F*-test and *t*-test are equivalent, and the latter can be used for directional alternatives. Hence for pairwise HOV comparisons (i.e., in the two-sample case), we use BF test as the usual *t*-test on the residuals from the median.

For the pooled LCDM distances by group and censored distances, there is an inherent spatial correlation between neighboring voxels which implies the dependence between LCDM distances (and hence the residuals from the medians) of the close-by voxels. Furthermore, LCDM distances (and hence the residuals from the medians) are significantly non-normal. Previously, it has been shown in [8] that the assumption violations for the parametric tests (ANOVA *F*-test and *t*-test) and for nonparametric tests (Kruskal-Wallis (KW) and WRS tests) have negligible effect on the empirical size and power performance of these tests. In this article, we also check the influence of assumption violations on the ANOVA of the residuals from the median, i.e., on BF HOV test.

The similarity in the morphometry of ROIs implies similarity of LCDM distances, which in turn implies similarity of the distributions of LCDM distances (hence similarity in the means, medians, and variances). That is, identical morphometry in the ROIs would imply identical distance distributions. But converse is not necessarily true in the sense that two ROIs might have similar distance distributions, but the corresponding morphometry could be very different. However, when the distance distributions are found to be different, it would logically imply different morphometry in the ROIs. For KW test, which is a nonparametric test, we test the equality of the distributions of the left pooled distances between $k$ groups; i.e., $H_o : F_1^L = F_2^L = \ldots = F_k^L$ where $F_i^L$ is the distribution function of the left pooled distances for group $i = 1, 2, \ldots, k$. The null hypothesis for the right distances is similar with $L$ being replaced with $R$. For ANOVA *F*-tests with or without HOV, we test the equality of the means of the left pooled distances between $k$ groups; i.e., $H_o : \mu_1^L = \mu_2^L = \ldots = \mu_k^L$ where $\mu_i^L$ is the mean of the left pooled distances for group $i = 1, 2, \ldots, k$. The null hypothesis for the right distances is similar with $L$ being replaced with $R$.

Observe that KW, ANOVA or WRS tests suggest shape and size differences when rejected, in particular the direction of the alternatives for the WRS test might indicate cortical thinning. Similarity of the morphometry of ROIs will cause similarity of LCDM distances, which would also imply similarity of the variances of LCDM distances. Variance of distances is suggestive of morphometric variation in ROIs. So similar shapes and sizes imply similar variances, but not vice versa. For example, cortical thinning might reduce the variation in LCDM distances, and the larger the spread in the boundary (surface) of ROI, the larger the variance of LCDM distances. In HOV analysis, we test the equality of the variances of the left pooled distances between $k$ groups; i.e.,

$$H_o : \text{Var}(D_1^L) = \text{Var}(D_2^L) = \ldots = \text{Var}(D_k^L) \tag{3}$$

where $\text{Var}(D_i^L)$ is the variance of the left pooled distances for group $i = 1, 2, \ldots, k$. The null hypothesis for the right distances is similar with $L$ replaced with $R$.

## 3 The Influence of Assumption Violations and Aggregation of Censored Distances on the HOV Tests: A Monte Carlo Study

The influence of the assumption violations due to the spatial correlation and nonnormality (i.e., non-Gaussianity) in the pooled LCDM distances has been shown to have negligible effect on the tests of location and distribution (such as ANOVA $F$-test, pairwise $t$-tests, KW test, WRS test, and KS test) [8]. In censoring LCDM distances, in addition to the above violations, we have the issue of accumulation/aggregation of the distances at each step. This aggregation of the distances have been shown not to substantially influence the tests and their sensitivity to the differences between the groups [20]. In this article, we assess the influence of the above problems on HOV tests. That is, we use an extensive Monte Carlo simulation study to determine the influence of violations of within sample independence and normality of pooled LCDM distances on the BF HOV test; and in censoring distances, in addition to these violations, we investigate the effect of distance accumulation at each censoring step on the HOV test. We employ the same Monte Carlo simulation setting of [8] in our data generation. For completeness, we replicate the distance generation procedure below which is shown to generate distances resembling those of LCDM distances from real subjects; i.e., capturing the true randomness in LCDM distances.

## 3.1 Simulation of Distances Resembling LCDM Distances

We choose the left VMPFC of HR subject 1 whose distances are denoted as $\mathfrak{D}_{21}^L = \{D_{21k}^L, k = 1, 2, \ldots, m_{21}^L\}$. We partition the range of distances into intervals $I_0 := [-1, 0.5]\, mm$, $I_1 := (0.5, 1.0]\, mm$, $I_2 := (1.0, 1.5]\, mm$, ..., and $I_{11} := (5.5, 6.0]\, mm$. Let $v_i^o$ be the number of distances within interval $I_i$, i.e., $v_i^o = |\mathfrak{D}_{21}^L \cap I_i|$, for $i = 0, 1, 2, \ldots, 11$. Then for $\mathfrak{D}_{21}^L$ we have $\vec{v}_o = (v_0^o, v_1^o, \ldots, v_{11}^o) =$ (2059, 1898, 1764, 1670, 1492, 1268, 814, 417, 142, 81, 61, 16). A possible Monte Carlo simulation to obtain LCDM-like distances can be performed as follows. We generate $n = 10000$ numbers with replacement in $\{0, 1, 2, \ldots, 11\}$ proportional to the above frequencies, $v_i^o$ (the choice of $n = 10000$ is due to the fact that the number of distances for left VMPFC of HR subject 1 is 11659). Then we generate as many uniform numbers in $(0, 1)$ (i.e., numbers from $\mathcal{U}(0, 1)$ distribution) for each $i \in \{0, 1, 2, \ldots, 11\}$ as $i$ occurs in the generated sample of 10000 numbers, and add these uniform numbers to $i$. Then we divide each distance by 2 to match the range of generated distances with $[0, 6.0]$ which is roughly the range of $\mathfrak{D}_{21}^L$. More specifically, we independently generate $n$ numbers from $\{0, 1, 2, \ldots, 11\}$ with the discrete probability mass function $P_o(N_j = i) = v_{p,i}^o = v_i^o / 11659$ for $i = 0, 1, \ldots, 11$ and $j = 1, 2, \ldots, n$. So, $P_o(N_j = i) = v_{p,i}^o$ where

$$(v_{p,0}^o, v_{p,1}^o, \ldots, v_{p,11}^o) = \vec{v}_p^o = (.177, .163, .151, .143, .126, .109, .070, .036, .012, .007, .005, .001).$$

Let $n_i$ be the frequency of $i$ among the $n$ generated numbers from $\{0, 1, 2, \ldots, 11\}$ with distribution $P_0$, for $i = 0, 1, \ldots, 11$. Hence $n = \sum_{i=0}^{11} n_i$. Then the set of simulated distances is

$$D_{mc} = \left\{ (J_s + U_s)/2 : J_s \stackrel{iid}{\sim} P_0 \text{ and } U_s \stackrel{iid}{\sim} \mathcal{U}(0,1) \text{ and } J_s \text{ and } U_s \text{ are independent for } s = 1, 2, \ldots, n \right\}.$$

See Figure 1 for histograms overlaid with the kernel density estimates of LCDM distances from HR subject 1 and 10000 simulated distances as described above. Notice that the histograms and the kernel density estimates are very similar, which indicates that distances generated by the above simulation procedure resembles distances from real-life VMPFCs.

## 3.2. Empirical Size Estimates and Size Curves
### 3.2.1. Multi-Sample Case

In the multi-sample case, our null hypothesis is HOV (i.e., equality of the variances) of LCDM distances, which follows from the equality of the distribution of LCDM distances for the groups. For simplicity, we consider $k = 3$ groups, extension of the below discussion to $k > 3$ groups is straightforward. Thus, for the null case, we generate three samples $\mathcal{X}, \mathcal{Y},$ and $\mathcal{Z}$ each of size $n_x, n_y,$ and $n_z$, respectively, as described above in Section 3.1 with the sample sizes for bins (stacks) being selected to be proportional to the frequencies $\vec{v}_o = (v_0^o, v_1^o, \ldots, v_{11}^o)$, i.e., distances are generated to be similar to the left VMPFC of HR subject 1. No generality is lost here, because distances for any other VMPFC can either be obtained by rescaling of the generated distances, or by modifying the frequencies in $\vec{v}_o$. In particular, sample $\mathcal{X}$ is generated as

$$D_{mc}^{\mathcal{X}} = \left\{ (J_s + U_s)/2 : J_s \stackrel{iid}{\sim} P_o \text{ and } U_s \stackrel{iid}{\sim} \mathcal{U}(0,1) \text{ and } J_s \text{ and } U_s \text{ are independent for } s = 1, 2, \ldots, n_x \right\}, \quad (4)$$

Samples $\mathcal{Y}$ and $\mathcal{Z}$ are generated similarly and denoted as $D_{mc}^{\mathcal{Y}}$ and $D_{mc}^{\mathcal{Z}}$, respectively.

We repeat this sample generation procedure $N_{mc} = 10000$ times and count the number of times the null hypothesis is rejected at $\alpha = 0.05$ level for BF test of HOV, KW test of distributional equality, and ANOVA $F$-tests (with and without HOV) of equality of mean distances, thus obtain the estimated significance levels under $H_o$. The estimated significance levels for various values of $n_x$, $n_y$, and $n_z$ are provided in Table 1, where $\hat{\alpha}_{BF}$ is the empirical size estimate for BF test, $\hat{\alpha}_{KW}$ is for KW test, $\hat{\alpha}_{F_1}$ is for ANOVA $F$-test with HOV, and $\hat{\alpha}_{F_2}$ is for ANOVA $F$-test without HOV; furthermore, $\hat{\alpha}_{KW,F_1}$ is the proportion of agreement between KW test and ANOVA with HOV, i.e., the number of times out of 10000 Monte Carlo replicates both KW test and ANOVA $F$-test with HOV simultaneously reject the null hypothesis. Similarly, $\hat{\alpha}_{KW,F_2}$ is the proportion of agreement between KW test and ANOVA $F$-test without HOV, and $\hat{\alpha}_{F_1,F_2}$ is the proportion of agreement between ANOVA $F$-tests with and without HOV, $\hat{\alpha}_{BF,KW}$ is the proportion of agreement between BF test and KW test, $\hat{\alpha}_{BF,F_1}$ is the proportion of agreement between BF and ANOVA $F$-test with HOV, and $\hat{\alpha}_{BF,F_2}$ is the proportion of agreement between BF test and ANOVA $F$-test without HOV. Using the asymptotic normality of the proportions, we test the equality of the empirical size estimates at 0.05 level, and compare the empirical sizes pairwise. We observe that all tests are at the desired level (i.e., empirical size estimates are not significantly different from the nominal level of 0.05; with $N_{mc} = 10000$, empirical size estimates within [.0464,.0536] are not significantly different from the nominal level of 0.05). Hence, if the distances are not that different; i.e., the frequency of distances for each bin and the distances for each bin are identically distributed for each group, the inherent spatial correlation does not seem to influence the significance levels. Moreover, the proportion of agreement between KW test and ANOVA $F$-tests with and without HOV are significantly less than 0.05 and also significantly less than the smaller of the empirical sizes in each pair (i.e., the proportion of agreement between KW test and ANOVA $F$-test with HOV is significantly smaller than the smaller empirical size of these tests). This implies KW test and ANOVA $F$-tests have significantly different rejection (and hence different acceptance) regions. This is in agreement with the fact that KW and ANOVA $F$-tests are actually testing different hypotheses; in fact, KW test is for distributional equality (based on ranks), while ANOVA $F$-tests are for equality of the means. On the other hand, the proportion of agreement between ANOVA $F$-tests, $\hat{\alpha}_{F_1,F_2}$, is neither signif-icantly smaller than 0.05, nor significantly smaller than the minimum of $\hat{\alpha}_{F_1}$ and $\hat{\alpha}_{F_2}$. Hence, $F_1$ and $F_2$ tests have about the same rejection (and acceptance) regions. That is, under the simulation of the null case, HOV is retained, hence ANOVA $F$-tests with or without HOV have the same empirical size performance, and moreover, $F_1$ and $F_2$ basically test the same hypotheses. In Table 1, we also observe that proportions of agreement with BF test and other tests (i.e., KW, $F_1$ and $F_2$ tests) are all significantly smaller than 0.05 (in fact, they are about 0.005), and also significantly smaller than the minimum of the empirical sizes in each pair. This suggests that the rejection and acceptance regions for BF test and all other tests are very different. In fact, BF test is testing equality of variances, while the others are tests of location or distribution (such as equality of means or rankings). Furthermore, the proportions of agreement between the tests of location is much higher compared to those of BF test with other tests, as the corresponding rejection or acceptance regions have different intersection levels. In particular, the common rejection region for tests of location is much larger than the common rejection region of BF test with a test of location or distribution.

For censoring, we use the pooled distances with $n_x = n_y = n_z = 10000$. For example for sample $\mathcal{X}$, we partition the range of generated distances into bins of size $\delta = 0.01$, then we have $\lfloor d_{\max}^{\mathcal{X}}/\delta \rfloor$ many bins where $d_{\max}^{\mathcal{X}}$ is the largest distance value in $D_{mc}^{\mathcal{X}}$. At $k^{th}$ censoring step, we only consider the distances less than or equal to $\gamma_{\delta,k} = k\delta$. These distances are denoted as

$$C_d^{\mathcal{X}}(k,\delta) := \left\{ d \in D_{mc}^{\mathcal{X}} \cap [0,k\delta] \right\} = \left\{ d \in D_{mc}^{\mathcal{X}} : d \leq k\delta \right\}.$$

Censored distances for samples $\mathcal{Y}$ and $\mathcal{Z}$ are obtained similarly from $D_{mc}^{\mathcal{Y}}$ and $D_{mc}^{\mathcal{Z}}$ and are denoted as $C_d^{\mathcal{Y}}(k,\delta)$ and $C_d^{\mathcal{Z}}(k,\delta)$, respectively. For brevity in notation, we write $\text{Var}\left(C_d^{\mathcal{X}}(k,\delta)\right)$ as $\text{Var}(X)$, $\text{Var}\left(C_d^{\mathcal{Y}}(k,\delta)\right)$ as $\text{Var}(Y)$, and $\text{Var}\left(C_d^{\mathcal{Z}}(k,\delta)\right)$ as $\text{Var}(Z)$. We repeat the same procedure $N_{mc} = 1000$ times. At each censoring step, we record the $p$-values for multi-group BF HOV test and KW test of distributional equality, and pairwise BF HOV tests and pairwise WRS tests. We also count the number of times the null hypothesis is rejected at $\alpha = 0.05$ level for these tests, thus obtain the empirical significance levels (i.e., sizes) under $H_o$. The average $p$-values and empirical size estimates together with 95% confidence bands for multi-group BF HOV test are plotted in Figure 2; and for pairwise BF HOV test for the one-sided alternatives $\text{Var}(X) < \text{Var}(Y)$ and $\text{Var}(X) > \text{Var}(Y)$, the empirical sizes are about 0.05 and average $p$-values are about 0.50 for all the tests considered hence are not presented.

### 3.2.2. Pairwise Size Comparisons
In the multi-sample case with three or more samples, the alternatives are general ones with no direction. However, if the multi-class omnibus tests (i.e., BF HOV test, and KW test, ANOVA $F$-tests for testing the distribution/location differences) are significant, then the next question of interest is which pair(s) of the samples exhibit differences and in which direction. To answer these questions, we can apply the two-sample versions of these tests, namely, BF test, WRS test, and Welch's $t$-test, as pairwise post-hoc tests. This will enable us to assess whether the methods/tests detect the specific direction of difference(s) between the samples which are present by construction; and if so, whether they detect these differences with high power.

***Remark 1:*** Two-sample tests can serve as a post-hoc testing procedure, after the multi-sample test being significant. If there are only two samples in the data set, then these tests are the two-sample versions of the multi-sample tests. As post-hoc tests, they may suffer from the type I error inflation because the two-sample tests will be performed conditioning on the significant result of the multi-sample tests when there are three or more classes. However, the conditional size and power estimates (conditional on the multi-sample test being significant) are only slightly different from unconditional ones. Hence, we only present the unconditional size and power estimates in this article. Furthermore, we omit the Holm's correction for pairwise comparisons between groups in our simulations (but in the application to the example data, Holm's correction is applied for pairwise group comparisons) for the same reason and to better compare the proportions of agreement between the two sample tests. ∎

We count the number of times the null hypothesis is rejected at $\alpha = 0.05$ for the post-hoc tests (BF test, WRS test, and $t$-test) and also Lilliefor's test of normality and KS test; thereby obtain the empirical size estimates for these tests. Although we have one type of alternative in the multi-sample case, for the two-sample case, except for Lilliefor's test, there are three types of alternative hypotheses possible: two-sided, left-sided, and right-sided alternatives. The estimated significance levels are provided in Table 2, where $\hat{\alpha}_{BF}$ is the empirical size estimate for BF test, $\hat{\alpha}_W$ is for WRS test, $\hat{\alpha}_t$ is for $t$-test, $\hat{\alpha}_{KS}$ is for KS test. Furthermore, $\hat{\alpha}_{W,t}$ is the proportion of agreement between WRS test and $t$-test, $\hat{\alpha}_{W,KS}$ is the proportion of agreement between WRS test and KS test, and $\hat{\alpha}_{t,KS}$ is the proportion of agreement between the $t$-test and KS test, $\hat{\alpha}_{BF,W}$ is the proportion of agreement between BF test and WRS tests, $\hat{\alpha}_{BF,t}$ is the proportion of agreement between BF test and the $t$-test, and $\hat{\alpha}_{BF,KS}$ is the proportion of agreement between BF test and KS test. We only present comparison of samples $\mathcal{Y}$ and $\mathcal{Z}$, since the sample size combinations for samples $\mathcal{X}$ and $\mathcal{Z}$ are included in $\mathcal{Y}$ vs $\mathcal{Z}$ comparisons. For $\mathcal{X}$ vs $\mathcal{Y}$ comparisons, the sample sizes (1000,1000) and (10000,10000) are included in $\mathcal{Y}$ vs $\mathcal{Z}$ comparisons, and (5000,5000) and (5000,7500) comparisons are omitted for brevity (as they would not provide anything new to the conclusions). Our samples are severely non-normal by construction, and normality is rejected for almost all samples generated, hence we omit the results of Lilliefor's test. We observe that under $H_o$, the empirical significance levels are about the desired level for all three types of alternatives, although KS test is slightly conservative. Hence, if the distances are not that different; i.e., the frequency of distances for each bin and the distances for each bin are identically distributed for each group, the inherent spatial correlation does not influence the significance levels for these tests. However, WRS, $t$-test, and KS tests are used for testing different null hypotheses, so their acceptance and rejection regions are significantly different for LCDM distances, since the proportion of agreement for each pair is significantly smaller than the minimum of the empirical size estimates for each pair of tests. Among these tests, the proportion of agreement between KS test and $t$-test is smallest. Moreover, we observe that the proportion of agreement of BF test with each of WRS, $t$, and KS tests is significantly smaller than 0.05 (in fact, the range of $\hat{\alpha}_{BF,W}$ is 0.0039 to 0.0055, the range of $\hat{\alpha}_{BF,t}$ is 0.0060 to 0.0088, and the range of $\hat{\alpha}_{BF,KS}$ is 0.0063 to 0.0091), and these proportions are much smaller than the agreement proportions for each pair of WRS, $t$, and KS tests.

### 3.3. Empirical Power Comparisons
### 3.3.1. Multi-Sample Power Comparisons of the Pooled Distances

For the alternative hypotheses in the multi-sample case, we again consider $k=3$ groups, namely, $\mathcal{X}, \mathcal{Y}$, and $\mathcal{Z}$. Let $\eta$ be a nonnegative integer and $\vec{v}_a(\eta) = \left(v_0^a(\eta), v_1^a(\eta), \ldots, v_{12}^a(\eta)\right)$ where $v_i^a(\eta)$ is the $i^{th}$ value after the entries $|v_i^o - \eta|$ are sorted in descending order for $i = 0,1,2,\ldots,11$ and $v_{12}(\eta) = 11659 - \sum_{i=0}^{11} |v_i^o - \eta|$. Then we set the probability mass function to

$$P_\eta(J=i) = v_{p,i}^a(\eta) = v_i^a(\eta) \Big/ \sum_{i=0}^{12} v_i^a(\eta).$$

Let $r \geq 1$ be a real number, then the set of simulated distances is

$$\left\{ (J_s + U_s)/2 : J_s \stackrel{iid}{\sim} P_\eta \text{ and } U_s \stackrel{iid}{\sim} \mathcal{U}(0,r) \text{ and } J_s \text{ and } U_s \text{ are independent for } s=1,2,\ldots,n \right\}.$$

Then for samples $\mathcal{X}, \mathcal{Y}$, and $\mathcal{Z}$, we set $r=r_x, \eta=\eta_x$, $r=r_y, \eta=\eta_y$, and $r=r_z, \eta=\eta_z$, respectively and take $n = n_x = n_y = n_z = 10000$. In our simulations, sample $\mathcal{X}$ distances are generated as in the null case with $r_x=1.0, \eta_x=0$; i.e., they are similar to distances of HR subject 1. So we have $v_{p,i}^a(\eta_x) = v_{p,i}^o$ for $i=0,1,\ldots,11$ and $v_{p,12}^a(\eta_x)=0$ where $v_{p,i}^o$ is defined in Section 3.1. Notice also that when $r_y = r_z = 1.0$ and $\eta_y = \eta_z = 0$, we obtain the null case of distributional equality between samples $\mathcal{X}, \mathcal{Y}$ and $\mathcal{Z}$. For practical purposes, we require $\eta_y$ and $\eta_z$ to be in $[0,1000]$ and $r_y$ and $r_z$ to be in $[0,2)$, although larger values would also be conceivable. The alternative cases we consider are

$$L1:(r_y,r_z,\eta_y,\eta_z)=(1.1,1.0,0,0); \qquad L2:(r_y,r_z,\eta_y,\eta_z)=(1.1,1.2,0,0); \qquad (5)$$
$$L3:(r_y,r_z,\eta_y,\eta_z)=(1.0,1.0,10,0); \qquad L4:(r_y,r_z,\eta_y,\eta_z)=(1.0,1.0,10,30).$$

See Figure 3 for the kernel density estimates of sample distances under the null case and various alternatives.

***Remark 2:*** Among the alternative parameters, $r>1$ tends to make the corresponding samples less variable compared to the null samples. More specifically, $r$ in $(1,2)$ would make a mild distance clustering for values around $(i+r)/2$ for $i=0,1,\ldots,11$ and does not increase the range and mean or medians of the distribution, but mostly changes the distribution and ranking of the distances and hence the locations of the samples. On the other hand, the parameters $\eta>0$ would make the distributions more skewed to the right, and also modify the underlying probability mass function. That is, it would change the distribution of the distances, and increase the variability, mean/median and ranges of the distances. We observe in our simulations that impact of $\eta>0$ on the variation of the samples is more severe compared to that on the distribution of the samples (see below). ∎

The sample generation procedure is repeated $N_{mc}=10000$ times. The empirical power estimates under $H_a$ are obtained as follows: We count the number of times the null hypothesis is rejected at $\alpha=0.05$ for BF HOV test, KW test of distributional equality, and ANOVA $F$-tests (with and without HOV) of equality of mean distances. The empirical power estimates are provided in Table 3, where $\hat{\beta}_{BF}$ is the empirical power estimate for BF test, $\hat{\beta}_{KW}$ is for KW test, $\hat{\beta}_{F_1}$ is for ANOVA $F$-test with HOV, and $\hat{\beta}_{F_2}$ is for ANOVA $F$-test without HOV. Note that as $n$ increases, the power estimates also increase under these alternative cases. Using the asymptotic normality of the empirical power estimates, we observe that under alternative cases $L1$ and $L2$, the variances of the distances are not that different, so we still have power estimates for BF test around .05 (i.e., in terms of variance differences, these alternatives are not different from the null case). However, under these alternatives, the distributions of the generated distances are different, hence the power estimates for the distribution and location tests (i.e., KW and ANOVA $F$-tests) are significantly larger than 0.05. In fact, the larger the $r_y-1.0$ and $r_z-1.0$, the higher the power estimates for KW and ANOVA $F$-tests. Under these alternatives, KW test tends to be more powerful than ANOVA $F$-tests, since such alternatives influence the distribution (hence ranking) of the distances, more than the means of the distances. Furthermore, under these alternatives, shape differences are more emphasized compared to size or scale differences; here "size" refers mostly to the distance with respect to the GM/WM surface. We also note that ANOVA $F$-tests (i.e., $F_1$ and $F_2$) have about the same power estimates.

Under alternative cases $L3$ and $L4$, the variances of the distances tend to differ. Hence as $\eta_y$ and $\eta_z$ deviate more from 0 in the positive direction, the power estimates for BF test increase, and so do the power estimates of KW and ANOVA $F$-tests. Under these types of alternatives, BF test tends to be the most powerful of the tests considered, and ANOVA $F$-tests tend to be more powerful than KW tests, since the right skewness (tail) of distances are more emphasized, which in turn implies that the differences in the variances and in the mean distances are emphasized more. Under these alternatives, both the size or scale and shape are different. If the GM voxels from the GM/WM surface are at different distances, BF test is the most sensitive to the differences in LCDM distances, as these alternatives suggest more variability in LCDM distances. Furthermore, ANOVA $F$-tests are more sensitive to the differences in LCDM distances compared to KW test as these alternatives suggest more variability in means compared to the rankings. We also note that both ANOVA $F$-tests have about the same power estimates, which suggests that ANOVA $F$-test tends to be robust to deviations from HOV. Therefore, based on our Monte Carlo analysis, we observe that the spatial correlation between distances has a mild influence on the results. That is, the results based on BF HOV test on multiple samples are still reliable, although the assumption of within sample independence and normality of the residuals are violated.

### 3.3.2 Pairwise Power Comparisons for Pooled Distances

For the alternative cases $L1-L4$, we determine which pairs of samples exhibit significant differences for the analysis of pooled distances. Note that in case $L1:(r_y,r_z,\eta_y,\eta_z)=(1.1,1.0,0,0)$, the comparison of $\mathcal{X}$ vs $\mathcal{Z}$ correspond to no difference, hence would be same as the size comparisons; and the same holds for case $L3:(r_y,r_z,\eta_y,\eta_z)=(1.0,1.0,10,0)$, hence in these cases we omit the comparison of $\mathcal{X}$ vs $\mathcal{Z}$. In $L1:(r_y,r_z,\eta_y,\eta_z)=(1.1,1.0,0,0)$ and $L2:(r_y,r_z,\eta_y,\eta_z)=(1.1,1.2,0,0)$, the comparison of $\mathcal{X}$ vs $\mathcal{Y}$ yield the same results, hence $L2$ is presented (for $L1$ and $L2$); the same holds for the comparison of $L3:(r_y,r_z,\eta_y,\eta_z)=(1.0,1.0,10,0)$ and $L4:(r_y,r_z,\eta_y,\eta_z)=(1.0,1.0,10,30)$, hence $L4$ is presented (for $L3$ and $L4$). In the alternative cases, we consider BF HOV test, WRS test, $t$-test as post-hoc tests for the pairwise analysis, and also Lilliefor's test of normality and KS test, and estimate the empirical powers for these tests. We do not present the power estimates for Lilliefor's test of normality, since our data is severely non-Gaussian by construction, and so we get power estimates of 1.00 under both null and alternative cases. The power estimates are provided in Table 4, where $\hat{\beta}_{BF}$ is the power estimate for BF test, $\hat{\beta}_W$ is for WRS test, $\hat{\beta}_t$ is for $t$-test, $\hat{\beta}_{KS}$ is for KS test. We only present the power estimates for the left-sided alternatives for $\mathcal{X}<\mathcal{Y}$, $\mathcal{X}<\mathcal{Z}$, and $\mathcal{Y}<\mathcal{Z}$, since by construction, $\mathcal{X}$ distances tend to be smaller than $\mathcal{Y}$ and $\mathcal{Z}$ distances, $\mathcal{Y}$ distances tend to be smaller than $\mathcal{Z}$ distances, and the power estimates in the reverse directions are virtually zero.

Under alternative case $L2$, the variances of the distances are not that different, so we still have power estimates for BF test around 0.05 (that is, in terms of HOV, these cases are not significantly different from the null hypothesis). But the distributions start to differ; so as $r_y$ and $r_z$ deviate further away from 1.0 in the positive direction, the power estimates for WRS, $t$-test, and KS tests tend to increase. Furthermore, as the sample size, $n$, increases, the power estimates for these tests increase as well. As in the multi-sample case, under these alternatives, WRS test is more powerful than $t$-test, since the ranking of the distances are affected more than the mean distances under these alternatives. But KS test has the highest power estimates for all sample sizes considered. Thus, for differences in shape rather than the thickness from the GM/WM surface, KS test and WRS test are more sensitive (with the former test being more sensitive) than $t$-test.

Under alternative case $L4$, the variances of the distances tend to differ; as $\eta_y$ and $\eta_z$ deviate further away from 0 in the positive direction, the power estimates for BF test increase, and so do the power estimates of WRS, $t$-test, and KS tests. Note that as $n$ increases, the power estimates also increase under each alternative case. Under these alternatives, $t$-test is more powerful than WRS test, since mean distances are more affected than the rankings under such alternatives. KS test has higher power estimates for larger deviations from the null case with larger sample sizes. Furthermore, BF test tends to have the highest power estimates under these alternatives. These alternatives imply that the distances of the GM voxels are at different scales, BF test has the best performance for small differences, while for large differences, KS and BF test have about the same performance that is better than the others.

Therefore, based on our Monte Carlo analysis, the spatial correlation between voxels (hence distances) has a mild influence, if any, on our results. That is, the results based on BF HOV test are still reliable, although assumptions of within sample independence and normality of the residuals are violated) and the results based on the other tests (WRS test, $t$-test, and KS tests) for two samples are still reliable, although the assumption of within sample independence is violated (and normality for the $t$-test is also violated). However, WRS test is more sensitive against the shape differences of GM of VMPFCs with similar distances from the GM/WM boundary and BF HOV test cannot detect such differences, since in this case, the variation of distances are not sufficiently different between the two groups. On the other hand, the $t$-test and KS test are more sensitive against the differences of GM tissue with different distances from the boundary compared to WRS test; we also notice that in this case, BF test has the best performance in detecting such deviations from the null case. That is, the variation of distances is more emphasized compared to the differences in central locations or ranking of the distances, hence the highest power for BF tests.

### 3.3.3 Pairwise Power Comparisons for Censored Distances

For the censoring distances, we consider the alternative hypothesis in which we generate sample $\mathcal{X}$ as in the null case. For sample $\mathcal{Y}$, we set $r_y = 1.2$ and $\eta_y = 0$ and for sample $\mathcal{Z}$, we set $r_z = 1.0$ and $\eta_z = 50$. So the alternative hypothesis we consider is

$$L5 : (r_y, r_z, \eta_y, \eta_z) = (1.2, 1.0, 0, 50) \quad (6)$$

and we use $n_x = n_y = n_z = 10000$. So, $P_{\eta_y}$ is the same as $P_{\eta_x}$ (which is the null distribution) and $P_{\eta_z}(N_j = i) = v^a_{p,i}(\eta_z)$ where

$$\left(v^a_{p,0}(\eta_z), v^a_{p,1}(\eta_z), \ldots, v^a_{p,12}(\eta_z)\right) = v^a_p(\eta_z) =$$
$$(.171, .158, .146, .138, .121, .104, .065, .051, .031, .008, .003, .003, .001).$$

Notice that sample $\mathcal{Y}$ is generated so that the rankings of distances are more different than those of sample $\mathcal{X}$ rather than the distances from the GM/WM surface. By construction, sample $\mathcal{Y}$ would have larger variance than sample $\mathcal{X}$, while sample $\mathcal{Z}$ would have larger variance compared to other samples. Furthermore, by construction, sample $\mathcal{Y}$ contains distances that are more accumulated at intervals [0.5,0.6], [1.0,1.1], …,[5.5,5.6] compared to sample $\mathcal{X}$. Therefore, at distances around these intervals, the censored distances for sample $\mathcal{X}$ tend to be smaller than censored distances for sample $\mathcal{Y}$ around $\gamma_{.01,k}$ for $k = 50, 100, \ldots, 550$ (i.e., around $k \times 0.01 = 0.5, 1.0, \ldots, 5.5$). Also, the variance of sample $\mathcal{Y}$ would be smaller than sample $\mathcal{Z}$ around these intervals provided censoring distances are less than 4.0. On the other hand, comparing $v^a_{p,i}(\eta_z)$ with $v^o_{p,i}$, we see that sample $\mathcal{Z}$ is more likely to have distances more than 4.0 compared to those of sample $\mathcal{X}$ while sample $\mathcal{X}$ is more likely to have distances less than 4.0 compared to those of sample $\mathcal{Z}$. Hence, we expect that for distances larger than 4.0, the censored distances for sample $\mathcal{X}$ tend to be smaller than censored distances for sample $\mathcal{Z}$ at $\gamma_{.01,k}$ for $k \geq 400$ (i.e., $\gamma_{.01,k} \geq 4.0$). The same trend is expected on variances of these samples. That is, the variance of sample $\mathcal{Z}$ would be larger than the variance of sample $\mathcal{X}$ for censoring distances larger than 4.0.

We repeat the sample generation procedure $N_{mc} = 10000$ times. We count the number of times the null hypothesis is rejected at $\alpha = 0.05$ for BF test, KW test, and pairwise WRS tests, thus obtain the empirical power estimates under $H_a$.

The average $p$-values together with 95% confidence bands versus censoring distance values for multi-group BF HOV test and for multi-group KW test are plotted in Figure 4. Observe that the differences in variance and distribution are at about the same censoring distances except for the distances between (2.5,4.0): There are significant differences between group variances at about $\gamma_{.01,k} = 0.5, 1.0, \ldots, 2.5$, and for distance values larger than 4.0. The significant differences at steps of 0.5 increments is because of the construction of sample $\mathcal{Y}$, and significant differences for distances larger than 4.0 are due to sample $\mathcal{Z}$. However, for censoring distances within $(2.5, 4.0)$, the samples seem to satisfy HOV, but they still tend to exhibit differences in distribution at about 2.5, 3.0, 3.5, and 4.0.

The average $p$-values together with 95% confidence bands versus censoring distance values for BF HOV test for the one-sided alternatives $\text{Var}(X) > \text{Var}(Y)$, $\text{Var}(X) < \text{Var}(Z)$, and $\text{Var}(Y) < \text{Var}(Z)$ and for WRS tests for the left-sided alternatives $X < Y$ (which means $\mathcal{X}$ values tend to be smaller than $\mathcal{Y}$ values), $X < Z$, and $Y < Z$ are plotted in Figure 5. Based on BF HOV test for $\text{Var}(X) > \text{Var}(Y)$ alternative, we observe that censored distances for sample $\mathcal{X}$ tend to have smaller variance than censored distances for sample $\mathcal{Y}$ around $\gamma_{.01,k}$ for $k = 50$ and larger variance for $k = 100, \ldots, 250$ (i.e., smaller variance around $\gamma_{.01,k} = 0.5$ and larger variance around $\gamma_{.01,k} = 1.0, \ldots, 2.5$). For censored distances larger than 3.0, $\text{Var}(X)$ and $\text{Var}(Y)$ are not significantly different from each other. Except around $\gamma_{.01,k} = 0.5$, by construction there is moderate clustering of sample $\mathcal{Y}$ distances around $\gamma_{.01,k} = 1.0, \ldots, 2.5$, which makes $\text{Var}(Y)$ significantly smaller than $\text{Var}(X)$. Based on WRS test for $X < Y$ alternative, we observe that censored distances for sample $\mathcal{X}$ tend to be smaller than censored distances for sample $\mathcal{Y}$ around $\gamma_{.01,k}$ for $k = 50, 100, \ldots, 350$ and $k \geq 400$ (i.e., around $\gamma_{.01,k} = 0.5, 1.0, \ldots, 3.5$ and at $\gamma_{.01,k} \geq 4.0$). For censored distances larger than 4.0, the proportions $v_{p,i}^a(\eta_x)$ and $v_{p,i}^a(\eta_y)$ are not large enough for samples $\mathcal{X}$ and $\mathcal{Y}$ to balance the accumulation of distances around 4.0, 4.5, 5.0, and 5.5 for sample $\mathcal{Y}$. Hence, censored distances for sample $\mathcal{Y}$ are significantly larger than those of sample $\mathcal{X}$ for $\gamma_{.01,k} \geq 3.5$.

Based on BF test for $\mathrm{Var}(X) < \mathrm{Var}(Z)$ alternative, we observe that censored distances for sample $\mathcal{X}$ tend to have smaller variance than the censored distances for sample $\mathcal{Z}$ at $\gamma_{.01,k}$ for $k \geq 400$ (i.e., at $\gamma_{.01,k} \geq 4.0$). Because, for censored distances larger than 4.0, the proportions have larger weights for sample $\mathcal{Z}$. Hence, $\mathrm{Var}(Z)$ is significantly larger than $\mathrm{Var}(X)$ for $\gamma_{.01,k} \geq 4.0$. Based on WRS test for $\mathcal{X} < \mathcal{Z}$ alternative, we observe that censored distances for sample $\mathcal{X}$ tend to be smaller than censored distances for sample $\mathcal{Z}$ at $\gamma_{.01,k}$ for $k \geq 400$ (i.e., at $\gamma_{.01,k} \geq 4.0$). Because for censored distances larger than 4.0, the proportions have larger weights for sample $\mathcal{Z}$. Hence, censored distances for sample $\mathcal{Z}$ are significantly larger than those of sample $\mathcal{X}$ for $\gamma_{.01,k} \geq 4.0$. Hence samples $\mathcal{X}$ and $\mathcal{Z}$ show the same trend in variance and distribution under this alternative (which does not occur in general).

Based on BF test for $\mathrm{Var}(Y) < \mathrm{Var}(Z)$ alternative, we observe that censored distances for sample $\mathcal{Y}$ tend to have smaller variance than censored distances for sample $\mathcal{Z}$ around $\gamma_{.01,k}$ for $k = 100,\ldots,250$ and $k \geq 400$ (i.e., around $\gamma_{.01,k} = 1.0,\ldots,2.5$ and $\gamma_{.01,k} \geq 4.0$). Furthermore, censored distances for sample $\mathcal{Y}$ tend to have larger variance than censored distances for sample $\mathcal{Z}$ around $\gamma_{.01,k}$ for $k = 50$ (i.e., around $\gamma_{.01,k} = 0.5$). Except around $\gamma_{.01,k} = 0.5$, by construction there is moderate clustering of sample $\mathcal{Y}$ distances around $\gamma_{.01,k} = 1.0,\ldots,2.5$, which makes $\mathrm{Var}(Y)$ significantly smaller than $\mathrm{Var}(Z)$. On the other hand, for censored distances larger than 4.0, the proportions have larger weights for sample $\mathcal{Z}$. Hence, $\mathrm{Var}(Z)$ is significantly larger than $\mathrm{Var}(Y)$ for $\gamma_{.01,k} \geq 4.0$. Based on WRS test for $Y > Z$ alternative, we observe that censored distances for sample $\mathcal{Y}$ tend to be larger than censored distances for sample $\mathcal{Z}$ around $\gamma_{.01,k}$ for $k = 50,100,\ldots,350$ (i.e., around $\gamma_{.01,k} = 0.5, 1.0,\ldots,3.5$). For censored distances larger than 4.0, the proportions are not large enough for sample $\mathcal{Z}$ to make its censored distances larger than those of sample $\mathcal{Y}$. Hence, censored distances for sample $\mathcal{Z}$ are not significantly different from those of sample $\mathcal{Y}$ for $\gamma_{.01,k} \geq 4.0$. This also occurs because the proportions have larger weights for distances less than 4.0, and any parameter affecting these distances have more influence in censored distance analysis. Moreover, for distances within (1.0, 3.0) variance differences and distributional differences are in the opposite direction. That is, for censoring distance in this interval, variance of sample $\mathcal{Y}$ is significantly smaller than variance of sample $\mathcal{Z}$, while sample $\mathcal{Z}$ distances tend to be larger than sample $\mathcal{Y}$ distances at the same censoring distance values.

***Remark 3:*** We omit the assessment of assumption violations and data aggregation on the censoring distance analysis via Monte Carlo simulations, because the results are similar to the pairwise comparisons after the multi-group analysis in the multi-sample case in Section 3.2.2. That is, we have the same conclusions of power comparisons in that section extended to the two-sample case; i.e., the assumption violations and data aggregation have negligible influence on the size and power of the tests under consideration, in particular, for BF HOV test. ∎

***Remark 4:*** **The Choice of the Reference Subject for Monte Carlo Simulations:** We have chosen left VMPFC of HR subject 1 as our baseline or reference tissue in our Monte Carlo simulations. This subject is actually a typical one and does not seem to be an outlier in VMPFC morphometry, see, e.g., the kernel density estimate of LCDM distances in Figure 13 (where the reference subject is indicated with a thicker solid line). In fact, the choice of the reference subject is not so relevant in the simulations, as in the null case we generate distances from the distance distribution of this subject, so they would all be similar (in distances) to each other satisfying the underlying assumption behind pooling. Deviations from the null case (i.e., the alternatives) are generated by modifying the parameters (i.e., entries in the frequency vector $\vec{v}_o$). Distances resembling other subjects can also be generated by the same approach. For example, to generate distances similar to subject 17 in the Ctrl group, we could have taken the frequency vector to be $\vec{v}_c =$

(2606, 2507, 2405, 2068, 1813, 1513, 1205, 822, 512, 291, 146, 62) and then could have used it as our reference subject. ∎

## 3.4 Comparison of the Tests for Normal and Skewed Data Sets
### 3.4.1. Normal Data

One of the main conclusions of this work is that HOV test is (and other tests are) valid and applicable in practice for LCDM distances, although some assumptions (such as within sample independence and normality) are violated. These tests can be much more powerful when the underlying assumptions are met by the data sets. In the presence of assumption violations, these tests might require very large samples to attain good power, but LCDM data sets tend to be extremely large by construction (as the voxel sizes are usually taken at mm or half mm resolution or adjusted accordingly based on the ROI considered) so sample size is not an issue for LCDM distances. However, we want to study and compare the size and power performance of the tests when the underlying assumptions are met, and thus consider first the normally distributed data sets that also satisfy within and between sample independence. We choose the same sample sizes used for the simulation of the LCDM-like distances in Sections 3.2 and 3.3.

With $k=3$ groups, we generate three samples $\mathcal{X}$, $\mathcal{Y}$, and $\mathcal{Z}$ each of size $n_x$, $n_y$, and $n_z$, respectively, each from a normal distribution. In particular, we generate sample $\mathcal{X}$ as

$$\left\{ X_s \stackrel{iid}{\sim} N(\mu_x, \sigma_x) \text{ for } s = 1, 2, \ldots, n_x \right\}. \quad (7)$$

Samples $\mathcal{Y}$ and $\mathcal{Z}$ are generated similarly. For the null case, we choose $\mu_x = \mu_y = \mu_z = 3.35$ and $\sigma_x = \sigma_y = \sigma_z = 2.28$; these choices are made so that the means and standard deviations of the normal data (approximately) match those of the data generated under the null case with $(r_y, r_z, \eta_y, \eta_z) = (1.0, 1.0, 0, 0)$ in Section 3.2.

For the alternatives, we use

$N1: \mu_x = \mu_z = 3.35, \mu_y = 3.39; \sigma_x = \sigma_y = \sigma_z = 2.28,$

$N2: \mu_x = 3.35, \mu_y = 3.39, \mu_z = 3.40; \sigma_x = \sigma_y = \sigma_z = 2.28,$

$N3: \mu_x = \mu_z = 3.35, \mu_y = 3.39; \sigma_x = \sigma_z = 2.28, \sigma_y = 2.33,$

$N4: \mu_x = 3.35, \mu_y = 3.39, \mu_z = 3.42; \sigma_x = 2.28, \sigma_y = 2.33, \sigma_z = 2.37,$

which correspond to cases $L1-L4$, respectively. That is, the means and standard deviations for the normal data in cases $N1-N4$ are chosen so that they (approximately) match those of the data from cases $L1-L4$, respectively.

Under N1 (resp. N2), the samples $\mathcal{X}$ and $\mathcal{Y}$ (resp. all samples) are different in mean but same in variance. Under N3, sample $\mathcal{Y}$ is different from the others in location and variance; and under N4, samples are all different both in mean and variance. The estimated significance levels based on $N_{mc} = 10000$ Monte Carlo replicates at $\alpha = 0.05$ level for the multi-class tests are presented in Table 5 where all the tests are about the desired level of .05; and the same holds for the pairwise tests, hence their size estimates are not presented. So, the methods behave as expected in terms of empirical size when all the assumptions are satisfied. The proportions of agreement between the multi-sample and pairwise tests have the similar trend as in Tables 1 and 2 (hence omitted again).

**Pooled Analysis:** For pooled data, we present the power estimates for the multi-group tests under $N1-N4$ in Table 6. Notice that $F_1$ and $F_2$ have very similar power for all cases. Under N1 and N2, BF test has power about the nominal level of the test (i.e., .05), and ANOVA $F$-tests tend to be more powerful compared to KW test for samples larger than 1000. On the other hand, under N3 and N4, BF test has much higher power compared to other tests, and ANOVA $F$-tests are slightly more powerful than KW test for samples larger than 1000. These results are in agreement with the fact than under N1 and N2, variances are equal, and means are different; while under N3 and N4, we have differences in mean and variance.

We present the power estimates for the pairwise tests under the alternative cases $N1 - N4$ in Table 7 where we only present $\mathcal{X} < \mathcal{Y}$, $\mathcal{X} < \mathcal{Z}$, and $\mathcal{Y} < \mathcal{Z}$ alternatives, as these are the only plausible alternatives by construction (except for N1 and N2, where variances are equal). Based on the same reasoning for presenting L2 and L4 only in Table 4, we only present cases N2 and N4 here. The pairwise tests are also performed so as to check whether the correct direction in the differences is detected for each comparison. As expected, under N2, power estimates of the BF HOV test is at about the nominal level of .05 since variances are equal by construction; $t$-test seems to be most powerful, then comes WRS test, and then KS test. So under N2, since all the assumptions (including equality of the variances) are met, the parametric test of location (i.e., $t$-test) is more powerful than the WRS and KS tests. However, under N4, BF test has the highest power; and for $\mathcal{X} < \mathcal{Y}$ alternative WRS, $t$-test, and KS tests have power estimates as in case N2, but for $\mathcal{X} < \mathcal{Z}$ and $\mathcal{Y} < \mathcal{Z}$ alternatives these tests are more powerful compared to the case N2; and as sample sizes increase, KS test has the highest power (since in this case, both location and scale are different, and WRS and $t$-tests are sensitive to location only). Furthermore, under N2 (resp. under N4), the power estimates for $\mathcal{X} < \mathcal{Z}$ is higher for WRS, $t$-test, and KS test (resp. all tests) compared to other alternative directions, as sample $\mathcal{Z}$ is much different in location (resp. different in location and scale) than other samples in this case.

**Censoring Analysis under N5:** In censoring, we truncate the data at the threshold values, hence the censored data becomes non-normal for most of the censoring values although the original data is normal. For example, we generate 1000 iid samples of size 10000 from $N(0,1)$ data, and perform censoring at $\delta = .01$ with lowest threshold being -2.3 and highest being 3. Lilliefor's test on these data implies that until about 2.5 standard deviations above the mean, the censored data is significantly non-normal, and for larger values censored data can be deemed to be (not different from) normally distributed. In censoring the direction of high power can depend on the censoring threshold. For example, if the standard deviations are same for two samples from two different normal distributions, the data with the smaller mean will have more variance for lower censoring values until a certain threshold is reached. We choose censoring increments as $\delta = .02$, $d_{\max} = 10.5$ and censoring starts at 0 (as in the LCDM distances) and consider the alternative case

$$N5: \mu_x = 3.35, \mu_y = 3.44, \mu_z = 3.47; \sigma_x = 2.28, \sigma_y = 2.26, \sigma_z = 2.40,$$

which is chosen so that the means and variances (approximately) match those in the alternative case L5. We provide the average $p$-values for multi-group comparisons as a function of the censoring values in Figure 6. Notice that BF test suggests that significant differences in variation occur for values larger than 2, and KW test suggests that significant differences occur for values larger than 8 (while mildly significant differences occur for values in (3,4)). The censoring plots for the pairwise comparisons are presented in Figure 7. BF test implies that variation of samples $\mathcal{X}$ and $\mathcal{Y}$ do not significantly differ at any censoring values (since by hypothesis, their variances are very close to each other), but $\mathrm{Var}(X) < \mathrm{Var}(Z)$ for values larger than 5, and $\mathrm{Var}(Y) < \mathrm{Var}(Z)$ for values larger than 1.5; WRS test indicates that censored $\mathcal{X}$ values tend to be smaller than censored $\mathcal{Y}$ values for censoring values larger than 5, censored $\mathcal{X}$ values tend to be smaller than censored $\mathcal{Z}$ values for values larger than 7.5, censored $\mathcal{Y}$ values tend to be larger than censored $\mathcal{Z}$ values for censoring values in (1,5).

**Censoring Analysis under N1-N4:** Under these cases, instead of plotting the average $p$-values against censoring values, we only provide the ranges of the data for which the power estimates are significantly larger than the nominal level of .05 (for $N_{mc} = 1000$ Monte Carlo replications, this corresponds to the threshold power value of .0620) together with the maximum power estimate, $\widehat{\beta}_{\max}$, in this range of data values. The power estimates for the multi-group analysis of the censored values are presented in Table 9. Although, this seems to over-summarize the results for each plot, it is chosen for the purposes of brevity in presentation. For example, Table 9 summarizes 12 figures of censoring plots in four rows. In the multi-group analysis of the censored values, we observe that KW and ANOVA $F$-tests are usually about the desired level, except for the slight liberalness around (.16, 2.30); on the other hand, BF HOV is at the desired level except for it is slightly liberal for values in (.76,1.22) and (2.50,3.12). Highest power estimates are attained under case N4 with BF test having much higher power than the other tests (as this alternative has the highest variance differences together with the location differences). However, under N1 and N2, BF test has very small power, which is only slightly larger than .05 and on a shorter range compared to KW and *F*-tests. This is as expected, since under N1 and N2, the variances are same, but the means are different. Moreover, under N3 and N4, KW and *F*-tests have high power, except around the one standard deviation above the mean (which might be due to data aggregation nullifying the differences in the ranking and location of the data).

The pairwise comparisons of the censored values are presented in Table 10.

$\mathcal{X}$ **vs** $\mathcal{Y}$ **:** We observe that under N1 and N2, $\mathcal{X}$ values tend to be smaller than $\mathcal{Y}$ values for all censoring values with WRS test (resp. *t*-test) with $\widehat{\beta}_{\max}$=.344 (resp. $\widehat{\beta}_{\max}$=.347), and variance of $\mathcal{X}$ values is mildly larger than that of $\mathcal{Y}$ values for values up to 8.76, with $\widehat{\beta}_{\max}$=.114. Under N3 and N4, $\mathcal{X}$ values tend to be smaller (resp. larger) than $\mathcal{Y}$ values with WRS test for values larger than 5.9 (resp. less than 5.36) with $\widehat{\beta}_{\max}$=.324 (resp. $\widehat{\beta}_{\max}$=.180), and with *t*-tests the trend is similar with value ranges being shifted about .5 unit to the right, and variance of $\mathcal{X}$ values is smaller than variance of $\mathcal{Y}$ values for all values with $\widehat{\beta}_{\max}$=.613.

$\mathcal{X}$ **vs** $\mathcal{Z}$ **:** Under N1, $\mathcal{X}$ and $\mathcal{Z}$ values exhibit almost no difference in distribution and variation, while under N2, $\mathcal{X}$ values tend to be smaller than $\mathcal{Z}$ values for all values with WRS test (resp. *t*-test) with $\widehat{\beta}_{\max}$=.431 (resp. $\widehat{\beta}_{\max}$=.444), and variance of $\mathcal{X}$ values is mildly larger than that of $\mathcal{Z}$ values for values up to 8.94, with $\widehat{\beta}_{\max}$=.137. Under N3, $\mathcal{X}$ and $\mathcal{Z}$ values exhibit almost no difference in distribution and variation, while under N4, $\mathcal{X}$ values tend to be smaller (resp. larger) than $\mathcal{Z}$ values with WRS test for values larger than 5.9 with $\widehat{\beta}_{\max}$=.64 (resp. smaller than 5.6 with $\widehat{\beta}_{\max}$=.39), and with *t*-tests the trend is similar with value ranges being shifted about .5-.7 units to the right, and variance of $\mathcal{X}$ values is larger than that of $\mathcal{Z}$ values for all values with $\widehat{\beta}_{\max}$=.971.

$\mathcal{Y}$ **vs** $\mathcal{Z}$ **:** Under N1, $\mathcal{Y}$ values tend to be larger than $\mathcal{Z}$ values for all values with WRS and *t*-tests with $\widehat{\beta}_{\max}$=.34, and variance of $\mathcal{Y}$ values is mildly smaller than that of $\mathcal{Z}$ values for values between .34 and 8.52 with $\widehat{\beta}_{\max}$=.116. On the other hand, under N2, $\mathcal{Y}$ values tend to be slightly smaller than $\mathcal{Z}$ values for all values with WRS and *t*-tests with $\widehat{\beta}_{\max}$=.080, and variance of $\mathcal{Y}$ and $\mathcal{Z}$ values are about the same. Under N3, $\mathcal{Y}$ values tend to be smaller (resp. larger) than $\mathcal{Z}$ values with WRS test for values up to 5.38 with $\widehat{\beta}_{\max}$=.206 (resp. larger than 5.9 with $\widehat{\beta}_{\max}$=.324), with *t*-tests the trend is similar with value ranges being shifted about .6 units to the right and variance of $\mathcal{Y}$ and $\mathcal{Z}$ values are about the same. Under N4, $\mathcal{Y}$ values tend to be smaller (resp. larger) than $\mathcal{Z}$ values with WRS test for values higher than 6.3 with $\widehat{\beta}_{\max}$=.195 (resp. less than 5.7 with $\widehat{\beta}_{\max}$=.174), with *t*-tests the trend is similar with value ranges being shifted about .6-.8 units to the right, and variance of $\mathcal{Y}$ values is smaller than that of $\mathcal{Z}$ values for all values with $\widehat{\beta}_{\max}$=.438.

In summary, when comparing two samples $A$ and $B$, from $N(\mu_a, \sigma_a)$ and $N(\mu_b, \sigma_b)$, respectively, the differences between censored data depends on both mean and variance differences. For example, if $\mu_a - 3\sigma_a < \mu_b - 3\sigma_b$, then censored A values will be smaller than censored B values for lower censoring thresholds, while the variance of censored A values will be larger than variance of censored B values for lower censoring thresholds.

### 3.4.2. Skewed Data Example: Exponential Distribution

The LCDM distances are skewed right by construction and hence the assumption of normality is inherently violated. Our simulations suggest that this violation has a negligible effect on the performance of the tests considered. However, to see how the methods perform for other skewed data sets which satisfy within and between sample independence, we consider exponentially distributed data sets. We choose the same sample sizes we used for the simulation of the LCDM-like distances in Section 3.2.

With $k=3$ groups, we generate three samples $\mathcal{X}, \mathcal{Y}$, and $\mathcal{Z}$ each of size $n_x, n_y$, and $n_z$, respectively, so that each one would be a random sample from an exponential distribution. In particular, we generate sample $\mathcal{X}$ as

$$D_{mc}^{\mathcal{X}} = \left\{ X_s \stackrel{iid}{\sim} \text{EXP}(\lambda_x) \text{ for } s=1,2,...,n_x \right\}. \tag{8}$$

Samples $\mathcal{Y}$ and $\mathcal{Z}$ are generated similarly from $\text{EXP}(\lambda_y)$ and $\text{EXP}(\lambda_z)$ with generated distances being denoted as $D_{mc}^{\mathcal{Y}}$ and $D_{mc}^{\mathcal{Z}}$, respectively. For the null case, we choose $\lambda_x = \lambda_y = \lambda_z = \lambda = 1.0$; this choice is made so that the range of the simulated data approximately matches that of the LCDM distances. That is, the range of the LCDM distances is (0,5.5), and $P(0 < X < 5.5) \approx .996$ for $X \sim \text{EXP}(1)$ data.

For the alternatives, we mimic the alternative types we used for the LCDM distances in Section 3.3. In particular, we first generate 1000000 data points from EXP(1) distribution, and divide the range of the distribution into 12 parts as $[0,.0.5), [.5,1.0),...,[5.0,\infty)$ and count the number of data points that fall in each interval (or bin) and round the numbers to the hundredth digit (so that the numbers are at the same scale as the vectors $\vec{v}_o$ and $\vec{v}_a(\eta)$ in Sections 3.2 and 3.3). Hence we obtain the vector of frequencies $\vec{v}_{\exp} = (v_{\exp,0}, v_{\exp,1},...,v_{\exp,11}) = (3930, 2385, 1449, 878, 534, 324, 196, 121, 73, 43, 26, 16)$. For the alternative hypotheses in the multi-sample case, we again consider $k=3$ groups, namely, $\mathcal{X}, \mathcal{Y}$, and $\mathcal{Z}$. Let $\eta$ be a nonnegative integer and $\vec{v}_{\exp}(\eta) = (v_0^{\exp}(\eta), v_1^{\exp}(\eta),...,v_{12}^{\exp}(\eta))$ where $v_i^{\exp}(\eta)$ is the $i^{th}$ value after the entries $|v_{\exp,i} - \eta|$ are sorted in descending order for $i = 0,1,2,...,11$ and $v_{12}^{\exp}(\eta) = 9975 - \sum_{i=0}^{11} |v_{\exp,i} - \eta|$. Then we set the probability mass functions to

$$P_{\exp,\eta}(J=i) = v_{p,i}^{\exp}(\eta) = v_i^{\exp}(\eta) \Big/ \sum_{i=0}^{12} v_i^{\exp}(\eta).$$

Let $r \geq 1$ be a real number, then the set of simulated distances is

$$\left\{ (J_s + U_s)/2 : J_s \stackrel{iid}{\sim} P_{\exp,\eta} \text{ and } U_s \stackrel{iid}{\sim} \mathcal{U}(0,r) \text{ and } J_s \text{ and } U_s \text{ are independent for } s=1,2,...,n \right\}. \tag{9}$$

For sample $\mathcal{X}$, we pick $r_x = 1.0, \eta_x = 0$, then we would have data from EXP(1) distribution. The empirical size estimates for the multi-group tests are provided in Table 5 for various values of $(n_x, n_y, n_z)$. We observe that the empirical size estimates are close to the nominal level of .05, but with some deviations for BF and $F$-tests, and no deviation for KW test. This is due to the fact that for exponential data normality is violated, and hence nonparametric test of KW test has better size performance compared to the parametric tests.

In the alternatives, we generate data based on this distribution with the parameters as in Section 3.3.1, that is, we choose

$$E1: (r_y, r_z, \eta_y, \eta_z) = (1.1, 1.0, 0, 0); \quad E2: (r_y, r_z, \eta_y, \eta_z) = (1.1, 1.2, 0, 0);$$
$$E3: (r_y, r_z, \eta_y, \eta_z) = (1.0, 1.0, 10, 0); \quad E4: (r_y, r_z, \eta_y, \eta_z) = (1.0, 1.0, 10, 30).$$

**Pooled Analysis:** The power estimates for the multi-group tests under $E1-E4$ are presented in Table 6. As before, $F_1$ and $F_2$ tests have very similar power for all cases. Under E1, BF test has power about the nominal level of .05 and under E2 the power is slightly larger than .05, as under E1 and E2, $r$ values (which affect the location/ranking) are larger than 1.0. Under E1 and E2, KW test has the highest power and then come ANOVA $F$-tests. From E1 to E2 power estimates increase, since $r_z$ increases from 1.0 to 1.2. Under E3 and E4, BF test has the highest power, then come the $F$-tests and then KW test, since under these cases, the parameter $\eta$ (which affect the scale more than location) is larger than 0. In particular, from E3 to E4, power estimates increase since $\eta_z$ increases from 0 to 30.

The power estimates for the pairwise tests under the alternative cases $E1-E4$ are presented in Table 8 where we only present the alternatives $\mathcal{X} < \mathcal{Y}$, $\mathcal{X} < \mathcal{Z}$, and $\mathcal{Y} < \mathcal{Z}$, as by construction these are the only sensible alternatives for power estimation (except for E1 and E2, where variances are equal). Based on the same reasoning for presenting only L2 and L4 in Table 4, we only present cases E2 and E4 here. Under E2, among the comparisons, tests have highest power for $\mathcal{X} < \mathcal{Z}$ alternative, and for each alternative, the ordering of the power estimates is KS > WRS > $t$ > BF. Under E4, BF test has the highest power and then comes the $t$-test; for smaller samples the power order is WRS > KS and for larger samples KS > WRS.

**Censoring Analysis under E5:** For the censoring of the exponential data, we consider the alternative

$$E5: (r_y, r_z, \eta_y, \eta_z) = (1.2, 1.0, 0, 50)$$

which has the same parameters as in L5 in Equation (6) (as in the alternative case for the LCDM-like distances). For censoring, we use $\delta = .01$ and the range of data from (0,5.5) for both null and alternative cases. The estimated significance levels for the multi-class tests under the null cases are at about the desired level for almost all distances and are not presented.

We provide the average $p$-values for multi-group comparisons as a function of the censoring values in Figure 8. Notice that BF test suggests that significant differences in variation occur for values in $(1.0, 1.4) \cup (1.8, 2.1) \cup (2.7, 3.1) \cup (5.45, 5.5)$, and KW test suggests that significant differences occur for values larger than 1.0. The censoring plots for the pairwise comparisons are presented in Figure 9. We observe that BF test implies that $\text{Var}(X) < \text{Var}(Y)$ for values in $(1.8, 2.1) \cup (2.6, 3.1) \cup (3.8, 4.0)$ and $\text{Var}(X) > \text{Var}(Y)$ for values in $(1.0, 1.4)$; WRS test indicates that censored $\mathcal{X}$ values tend to be smaller than censored $\mathcal{Y}$ values for censoring values larger than 1.0. Moreover, BF test implies that $\text{Var}(X) < \text{Var}(Z)$ for values in $(5.3, 5.5)$, and $\mathcal{X}$ and $\mathcal{Z}$ values are not significantly different from each other at any censoring value. BF test implies that $\text{Var}(Y) < \text{Var}(Z)$ for values in $(1.8, 2.1) \cup (2.6, 3.1) \cup (5.45, 5.5)$ and $\text{Var}(Y) > \text{Var}(Z)$ for values in $(1.0, 1.4)$; WRS test indicates that censored $\mathcal{Y}$ values tend to be larger than censored $\mathcal{Z}$ values for censoring values larger than 1.0.

**Censoring Analysis under E1-E4:** For the multi-group and pairwise comparisons of the censored distances, we observe that all the tests are about the desired level under the null case, and hence size estimates are not presented. Moreover, we do not present the summary table of ranges of significant differences and $\widehat{\beta}_{\max}$ values, but list the significant findings below. Under E1 and E2, KW test has higher power (than .05), that is, KW is indicates significant differences for values larger than 1.0 with $\widehat{\beta}_{\max}=1.00$; under E3, KW indicates only slightly significant differences for the ranges (2.5,3.5) and (4.57, 5.5) with $\widehat{\beta}_{\max}=.073$; and under E4, the trend is as in E3 with the differences occurring for values larger than 2.43 with $\widehat{\beta}_{\max}=.099$. $F$-test has similar performance for the censoring distances with slightly higher $\widehat{\beta}_{\max}$ values. On the other hand, BF HOV test indicates significant variance differences for values larger than 1.01 under E1 with $\widehat{\beta}_{\max}=1.0$; under E3, BF test shows no significant difference for almost all the values, while under E4, it is slightly significant for values larger than 4.45 with $\widehat{\beta}_{\max}=.094$. BF test shows a very interesting behavior under E2, where the ranges of significant variance differences is $(1.50, 1.60) \cup (2.22, 2.37) \cup (3.09, 3.17) \cup (4.06, 4.54)$ with $\widehat{\beta}_{\max}=1.0$.

For the pairwise comparison of the censored distances under the alternatives, performance of WRS and $t$-tests are similar, hence we only present the results for WRS test.

$\mathcal{X}$ **vs** $\mathcal{Y}$: WRS indicates $\mathcal{X} < \mathcal{Y}$ for values larger than 1.01 with $\widehat{\beta}_{\max}=1.0$ under cases E1 and E2, and has no power for $\mathcal{X} <$ Y under cases E3 and E4. On the other hand, WRS has no power for $\mathcal{X} > \mathcal{Y}$ direction for cases E1 and E2 (in fact, virtually zero power for values larger than 1.01), and under cases E3 and E4, WRS test indicates $\mathcal{X} > \mathcal{Y}$ for values larger than 2.27 with $\widehat{\beta}_{\max}=.104$. BF HOV test indicates that variance of $\mathcal{X}$ is less than variance of $\mathcal{Y}$ for values in $(1.01, 1.52) \cup (2.08, 2.27) \cup (5.09, 5.46)$ with $\widehat{\beta}_{\max}=1.0$ under cases E1 and E2; and variance of $\mathcal{X}$ is larger than variance of $\mathcal{Y}$ for values in $(1.58, 2.06) \cup (2.31, 4.07) \cup (4.35, 5.05)$ with $\widehat{\beta}_{\max}=.873$ under cases E1 and E2, and for values larger than 2.93 with $\widehat{\beta}_{\max}=.102$ under case E3.

$\mathcal{X}$ **vs** $\mathcal{Z}$ : WRS test indicates that $\mathcal{X} < \mathcal{Z}$ for values larger than 1.01 with $\widehat{\beta}_{\max}=1.0$ under case E2; and $\mathcal{X} > \mathcal{Z}$ for values larger than 2.42 with $\widehat{\beta}_{\max}=.169$; and $\mathcal{X}$ and $\mathcal{Z}$ are not significantly different under other cases. BF HOV test indicates that variance of $\mathcal{X}$ is less than variance of $\mathcal{Z}$ for values in $(1.01, 1.58) \cup (2.16, 2.31) \cup (4.19, 4.27) \cup (5.15, 5.5)$ with $\widehat{\beta}_{\max}=1.0$ under case E2 and for values in $(.78,.84)$ with $\widehat{\beta}_{\max}=.070$ under case E4; and variance of $\mathcal{X}$ is larger than variance of $\mathcal{Z}$ for values in $(1.60, 2.14) \cup (2.34, 3.19) \cup (3.22, 4.16) \cup (4.46, 5.12)$ with $\widehat{\beta}_{\max}=.999$ under case E2, and for values in (2.25,5.46) with $\widehat{\beta}_{\max}=.184$ under case E4.

$\mathcal{Y}$ **vs** $\mathcal{Z}$ : WRS test indicates that $\mathcal{Y} < \mathcal{Z}$ for values larger than 1.01 with $\widehat{\beta}_{\max}=1.0$ under case E2; and $\mathcal{Y} > \mathcal{Z}$ for values larger than 2.65 with $\widehat{\beta}_{\max}=.120$; and $\mathcal{Y}$ and $\mathcal{Z}$ are not significantly different under other cases. BF HOV test indicates that variance of $\mathcal{Y}$ is less than variance of $\mathcal{Z}$ for values in  with $\widehat{\beta}_{\max}=.874$ under case E1, for values in $(1.11, 1.63) \cup (2.19, 2.34) \cup (3.20, 3.26) \cup (4.20, 4.36) \cup$ $(5.18, 5.50)$ with $\widehat{\beta}_{\max}=.998$ under case E2, for values larger than 2.48 with $\widehat{\beta}_{\max}=.107$ under case E3, and for values larger than 5.44 with $\widehat{\beta}_{\max}=.071$ under case E4; and variance of $\mathcal{Y}$ is larger than variance of $\mathcal{Z}$ for values in $(1.01, 1.51) \cup (2.08, 2.26) \cup (4.10, 4.18) \cup (5.08, 5.44)$ with $\widehat{\beta}_{\max}=.999$ under case E1, for values in $(1.67, 2.17) \cup (2.34, 3.18) \cup (3.37, 4.17) \cup (4.56, 5.16)$ with $\widehat{\beta}_{\max}=.808$ under case E2, and for values in (2.58,5.31) with $\widehat{\beta}_{\max}=.135$ under case E4.

### 3.5. Simultaneous Inference for HOV of Censored Distances

The HOV analysis of censored distances has been performed pointwise at each censoring step so far. Our Monte Carlo simulation study indicates that the censoring analysis correctly provides the ranges of distances at which there are significant differences between the groups. However, censored distance analysis is a type of simultaneous inference when all the threshold values are considered together (i.e. when inference is performed for all censored distances), and it is methodologically more reliable to correct for the multiple testing procedure (in particular, for our simulations, we have tests conducted at about 600 censoring threshold values). In statistical literature, there are many correction procedures for multiple testing. The most common ones are Bonferroni, Holm, Šidák corrections, Tukey procedure, Hochberg's step-up procedure [51], and more recently introduced ones such as Benjamini-Hochberg (BH) [41] and Benjamini-Yekutieli (BY) [52] corrections. Among these procedures, Bonferroni method is the most conservative (but perhaps most well-known) one and does not require independence of the tests, Šidák's procedure is more powerful than Bonferroni, but requires independence. Also, Holm's method is more powerful than Bonferroni, and does not require independence. Tukey's method is valid for pairwise comparisons only, hence not applicable for our censored distance analysis. BH procedure is usually about the nominal size for independent tests and in some types of dependence, and BY procedure is more conservative than BH but works for dependent tests as well. BY procedure employed in this article is the version suggested for dependent tests in [52]. BH and BY procedures are designed to control the prespecified false discovery rate (FDR) value, while the other procedures are designed to control for family-wise error rate (FWER). The test statistics (hence the $p$-values) are not independent in our censored distance analysis, hence Šidák's procedure and Hochberg's step-up procedure are not appropriate for our analysis.

    Benjamini and Yekutieli demonstrated that BH correction is also appropriate for test statistics satisfying positive regression dependency. Our experience suggests that the test statistics for HOV of censored distances satisfy a form of positive association (conditionally), and thus satisfies the positive regression dependency. Considering all these, we only use Bonferroni correction (as it is one of the most well-known and common procedures in multiple testing), Holm's correction, BH and BY procedure for adjusting $p$-values in our simultaneous inference.

    We perform these corrections on the $p$-values for multi-group BF HOV test provided in Figure 2 which were based on the simulated data under $H_o$. All corrected $p$-values except BH corrected ones are

virtually 1 at each censoring threshold value and BH corrected $p$-values are about .80 (hence not presented). Thus, the procedures except BH procedure are extremely conservative in testing HOV of censored distances. The trend is similar for the corrected versions of the $p$-values of pairwise tests provided in Figure 5, hence are not presented. We also perform corrections for the $p$-values for BF HOV test under the alternatives. In Figure 10, we present the corrected versions of the $p$-values for multi-group BF test in Figure 4 which are generated under the alternative case L5 in Equation (6). Notice that Bonferroni and Holm corrections do not detect any deviation from HOV for distances less than 4.0 (as they are extremely conservative). On the other hand, BY corrected version captures the general trend, but misses most of the distance ranges of significant differences and BH corrected version has the similar trend and catches almost all the distance ranges of significant differences. The same trend in the corrected $p$-values is observed for the pairwise BF tests under the alternatives, hence are not presented. Our Monte Carlo simulations suggest that correction for multiple testing for HOV analysis of censored distances seems not to be a crucial procedure, as without a correction BF test captures the distance ranges of significant differences correctly with almost no false discovery. However, for our simultaneous inference for ranges of distances to be more reliable, we recommend the use of BH correction for HOV tests of censored distances. This recommendation is based on our Monte Carlo analysis and some theoretical justification for (conditional) positive association of the tests.

## 4. Case Study: The HOV and Location/Distribution Analyses of LCDM Distances of VMPFCs

In this section, we apply our methodology on our example illustrative data set; that is, we analyze the morphometric variability of GM in VMPFCs in subjects with MDD, subjects with HR of MDD, and healthy subjects by the HOV and other tests on the LCDM distances. One of the crucial underlying assumptions behind our analysis is that the LCDM distances for left (resp. right) VMPFC of the subjects from the same diagnostic group have similar morphometry, hence the distances from the left (resp. right) VMPFCs for subjects in the same diagnostic group come from the same distribution. We can check validity of this assumption in various ways. For example, we can compare each subject's LCDM distances with respect to each diagnostic group by a test, say WRS test, and store the test statistic for each subject, and compare these test statistics for the subjects between the diagnostic groups. To make this approach more concrete, for example, subject $i$ in HR group is compared with the pooled HR group (subject $i$ excluded), and the same subject is compared with the pooled MDD distances and pooled Ctrl distances. We apply the same procedure to each subject in each group. For illustrative purposes, we present only the WRS test statistics for the HR subjects in Figure 11, where we observe that HR subjects are more similar to other HR subjects, and less similar to MDD subjects and least similar to the Ctrl subjects for both left and right VMPFCs. Moreover, one can also use exploratory means to visually determine whether the subjects in each diagnostic group are similar in morphometry to each other. Along this line, we plot the kernel density estimates of LCDM distances of left (resp. right) VMPFCs for the subjects in each diagnostic group at a separate plot in Figure 12. Notice that the subjects in each diagnostic group have VMPFCs similar in morphometry, except possibly a few (two to three) morphometric outliers for each diagnostic group. The methodology proposed in this article would very likely provide more reliable results if the outliers were detected and handled properly (e.g., they are excluded from further analysis or distances could be measured again for better accuracy); however, the issue of (morphometric) outlier detection by the use of LCDM distances is not handled in the current article and is a topic of ongoing research.

The sample sizes, means, medians, and the standard deviations of the LCDM distances for each group and overall are presented in Table 11. Notice that the order of the groups are same in mean, median, and standard deviation of LCDMs with HR < MDD < Ctrl for both left and right VMPFCs. This is suggestive of shrinkage in VMPFC due to MDD or being at HR for MDD, and also morphometric variability seems to be reduced related to MDD or being at HR for MDD.

### 4.1. The HOV Analysis of Pooled LCDM Distances of VMPFCs

We consider the differences in both location and spread for the pooled distances (by group). The histograms and the kernel density estimates (overlaid on the histograms) of the pooled distances for the left and right VMPFCs are presented in Figure 13. The left and right pooled distances for each group are significantly non-normal with $p < .0001$ based on Lilliefor's test of normality (see, e.g., [38]), due to heavy right skewness of the densities. The HOV is rejected with $p<.0001$ based on BF test. Hence we perform pairwise HOV comparisons to determine which pairs violate HOV. See Table 12 for the corresponding $p$-values for pairwise comparisons adjusted by Holm's correction method. The order of the variances is HR < MDD < Ctrl for both left and right VMPFCs with $p<.0001$ for all six possible comparisons. This implies that the morphometric variation reduces in left and right VMPFCs due to having MDD or being HR compared to Ctrl subjects and is smallest for the HR subjects for both left and right VMPFCs. However, most of the reduction in the morphometric variation is not due to shape but size; because as the VMPFCs shrink in size, the LCDM distances tend to have less variation and this size shrinkage might override or dominate the possible increases in shape variation (if exists) due to depression or being at HR.

The equality of the distributions of the distances of left VMPFCs is rejected with KW and ANOVA $F$-tests ($p<.0001$ for all tests). Likewise for right VMPFC distances ($p<.0001$ for all tests). Hence, we perform pairwise comparisons by WRS test and $t$-test for left (and right) VMPFC distances, using Holm's correction for multiple comparisons. The $p$-values adjusted by Holm's correction method for the simultaneous pairwise comparisons for left and right VMPFC distances are presented in Table 12. Observe that, with WRS test, MDD-left and HR-left distances are not significantly different and so are MDD-right and HR-right distances ($p=.6044$ for former, $p=.1552$ for latter), while both tend to be significantly less than Ctrl-left distances ($p<.0001$ for all comparisons). Likewise for right distances ($p<.0001$ for all comparisons). Observe also that WRS test and $t$-test (for location) and BF test (for variances) yield significant results with the same ordering between groups (i.e., HR < MDD < Ctrl), which might be due to cortical thinning among other factors. In Table 2, we also present the $p$-values for KS test. Notice that $p$-values for the one-sided tests are not complementary of each other (i.e., the sum of left- and right-sided alternatives do not add up to 1), since KS test declares significance for the maximum difference between the two samples, say A and B, and at one distance value sample A can have cumulative distribution function (cdf) larger than cdf of sample B, and at another one sample B can have cdf larger than cdf of sample A. This behavior occurs for comparing MDD vs HR left distances, MDD vs Ctrl right distances, and HR vs Ctrl right distances. On the other hand, cdf of each of MDD and HR left distances is significantly larger than cdf of Ctrl left distances, and cdf of HR right distances is significantly larger than cdf of MDD right distances. If cdf of a sample, say sample A, is larger than cdf of sample B, then it is more likely for sample A to have smaller values than sample B, which might imply thinning in the context of VMPFCs.

## 4.2. The HOV Analysis of the Censored LCDM Distances of VMPFCs

Recall that at each censoring distance, $\gamma_{\delta,k}$, we have the distance values in $\left[0.5, \gamma_{\delta,k}\right]$ mm. These censored distances convey shape/size information at the specified $\gamma_{\delta,k}$ value, i.e., at distance of $\gamma_{\delta,k}$ or less from the GM/WM surface. We only consider the comparisons for $\gamma_{\delta,k} \in [0.5, 5.5]$ mm, due to the confounding effect of negative distances. That is, the influence of negative distances makes the comparison for small censoring distance values unreliable, and this confounding influence becomes negligible for sufficiently large $\gamma_{\delta,k}$. Furthermore, at each censoring step $k$, the censored distances are severely non-normal. We perform Lilliefor's test of normality for each group at each censoring step $k$ for both left and right VMPFCs. The maximum $p$-value for MDD left censoring volumes among all $p$-values (based on Lilliefor's test) for censoring steps is $p_{L,1}^{\max} = 8.6 \times 10^{-18}$. Similarly, $p_{L,2}^{\max} = 1.1 \times 10^{-11}$, $p_{L,3}^{\max} = 1.9 \times 10^{-35}$, $p_{R,1}^{\max} = 9.1 \times 10^{-16}$, $p_{R,1}^{\max} = 1.3 \times 10^{-16}$, and $p_{R,1}^{\max} = 4.8 \times 10^{-27}$. Notice that all $p$-values are virtually zero, suggesting severe non-Gaussianity for censored distances at all censoring steps.

We also perform the tests of homogeneity of variance (HOV) of censored distances. The variance of censored distances is a measure of spread of censored distances, but since at each censoring distance value, we restrict the spread in the normal direction from the surface, it measures more of variation size in the parallel direction (width) and shape. At each step $k$, we perform the multi-group BF HOV test, and store the associated $p$-value. The corresponding null hypothesis for left censored distances is

$$H_o : \text{Var}\left(C_{d,1}^L(k,\delta)\right) = \text{Var}\left(C_{d,2}^L(k,\delta)\right) = \text{Var}\left(C_{d,3}^L(k,\delta)\right)$$

where $\text{Var}\left(C_{d,i}^L(k,\delta)\right)$ is the variance of left censored distances for group $i=1,2,3$. The null hypothesis for right censored distances is similarly defined.

The alternative has no direction when there are three or more groups. See Figure 14 for the $p$-values for multi-group HOV and KW tests (the results based on ANOVA $F$-tests are similar to the results of KW test, hence are omitted). Observe that multi-group HOV is rejected at about the same censoring distance value for both left and right VMPFCs. For left VMPFCs, there are variance differences in distances for censoring distance values 2.20 *mm* and higher, while for right VMPFCs, variance differences occur for censoring distance values 2.50 *mm* and higher. Based on KW test; we observe that the differences between distributions (means) of left and right censored distances start to occur at about the same $\gamma_{\delta,k}$ value. The distributions and means of the distances are significantly different for $\gamma_{\delta,k}$ values of 2.00 *mm* or larger for left VMPFCs, and 2.20 *mm* or larger for right VMPFCs. That is, differences for right VMPFCs start to occur at a slightly larger distance from the GM/WM surface.

To find out which pairs of groups manifest variance differences in censored distances, we perform pairwise BF HOV test. The null hypothesis for left censored distances is

$$H_o: \text{Var}\left(C_{d,1}^L(k,\delta)\right) = \text{Var}\left(C_{d,2}^L(k,\delta)\right) \text{ and } \text{Var}\left(C_{d,1}^L(k,\delta)\right) = \text{Var}\left(C_{d,3}^L(k,\delta)\right)$$
$$\text{and } \text{Var}\left(C_{d,2}^L(k,\delta)\right) = \text{Var}\left(C_{d,3}^L(k,\delta)\right).$$

For each pair, we conduct BF HOV test for both less-than and greater-than alternatives. The less-than alternative for the left censored distances is

$$H_a: \text{Var}\left(C_{d,1}^L(k,\delta)\right) < \text{Var}\left(C_{d,2}^L(k,\delta)\right) \text{ and } \text{Var}\left(C_{d,1}^L(k,\delta)\right) < \text{Var}\left(C_{d,3}^L(k,\delta)\right)$$
$$\text{and } \text{Var}\left(C_{d,2}^L(k,\delta)\right) < \text{Var}\left(C_{d,3}^L(k,\delta)\right).$$

The null and alternative hypotheses for right censored distances are similar. Then we adjust the $p$-values for pairwise HOV tests by Holm's correction method for both alternatives and plot the $p$-values against the censoring distance values. However, we omit the $p$-values corrected for multiple testing, because BH correction is the recommended method that works in our case and the results after BH correction are almost identical to the uncorrected ones. See Figure 15 for the $p$-values for the pairwise HOV test and pairwise WRS test for left VMPFCs. Observe that for left censored distances, the variance of MDD group is significantly less than Ctrl group for censoring distance values 2.3 *mm* and higher; the variance of HR group is significantly less than Ctrl group for censoring distance values 2.8 *mm* and higher; and the variance of MDD group is less (resp. greater) than HR group for $\gamma_{\delta,k} \in [2.00, 3.00]$ *mm* (resp. $\gamma_{\delta,k} > 3.60$ *mm*). Based on the plots of the one-sided $p$-values, we see that MDD left censored distances tend to be significantly less than Ctrl left censored distances for $\gamma_d(k,\delta)$ values of 2 *mm* and higher. That is, at distance values of 2 *mm* or larger from the GM/WM surface, it is more likely for a voxel to be in the exterior of GM of MDD left VMPFC compared to that of Ctrl left VMPFC. In other words, there are fewer GM voxels in left VMPFC of MDD group at distance values of 2 *mm* and higher compared to left VMPFC of Ctrl group. HR left censored distances are significantly smaller for $\gamma_d(k,\delta)$ values of 2.8 *mm* and higher compared to Ctrl left censored distances. The interpretation is as above. On the other hand, MDD left censored distances are significantly less than HR left censored distances for $\gamma_d(k,\delta)$ values between 2.2 and 3.2 *mm*, and larger than HR left censored distances for $\gamma_d(k,\delta)$ values between 4.5 *mm* and higher. Hence, there are fewer GM voxels in MDD left VMPFCs at distance values between 2.2 and 3.2 *mm*, and more GM voxels at distance values larger than 4.5 *mm* compared to HR left VMPFCs. Notice also the different results for MDD and HR left distance comparisons: MDD left distances tend to be larger than HR distances for distances at 5.0 *mm* or larger, while mean distance for MDD left VMPFCs is significantly larger than that of HR left VMPFCs at distances 4.5 *mm* or larger.

See Figure 16 for the $p$-values for the pairwise HOV test for right VMPFCs. For right censored distances, the variance of MDD group is significantly less than Ctrl group for censoring distance values 2.8 *mm* and higher; the variance of HR group is significantly less than Ctrl group for censoring distance values 2.4 *mm* and higher; and the variance of MDD group is greater than Ctrl group for $\gamma_d(k,\delta) > 2.6$ *mm*.

MDD right censored distances are significantly less than Ctrl right censored distances for $\gamma_d(k,\delta)$ values of 2.6 *mm* and higher. That is, at distance values of 2.6 *mm* or larger from the GM/WM surface, it is more likely for a voxel to be in the exterior of GM of MDD right VMPFC compared to that of Ctrl right VMPFC. In other words, there are fewer GM voxels in right VMPFC of MDD group at distance values of 2.6 *mm* and higher compared to right VMPFC of Ctrl group. HR right censored distances are significantly smaller for $\gamma_d(k,\delta)$ values of 2.2 *mm* and higher compared to Ctrl right censored distances. The interpretation is as above. On the other hand, MDD right censored distances are significantly larger than HR right censored distances for $\gamma_d(k,\delta)$ values of 2.2 *mm* or larger. Hence, there are more GM voxels in MDD right VMPFC at distance values 2.2 *mm* or larger compared to HR right VMPFCs.

**Remark 5: Summary of the VMPFC Comparisons:** Based on our HOV analysis on the LCDM distances of VMPFCs, we observe that there are variance differences in distances for censoring distance values 2.20 *mm* and higher for left VMPFCs, while variance differences occur for censoring distance values 2.50 *mm* and higher for right VMPFCs. Moreover, pairwise BF HOV test on the censored distances indicate that for left censored distances, the variance of MDD group is significantly less than Ctrl group for censoring distance values 2.3 *mm* and higher; the variance of HR group is significantly less than Ctrl group for censoring distance values 2.8 *mm* and higher; and the variance of MDD group is less (resp. greater) than HR group for $\gamma_{\delta,k} \in [2.00, 3.00]$ *mm* (resp. $\gamma_{\delta,k} > 3.60$ *mm*). If the variance is larger for a range of censored distances, it suggests more variation in the LCDM distances and hence more variation in morphometry for that range of distances in the group in question. For example, for the MDD left VMPFCs, the morphometric variability is less than the Ctrl group for distances 2.2 *mm* and higher from the GM/WM surface. For right censored distances, the variance of MDD group is significantly less than Ctrl group for censoring distance values 2.8 *mm* and higher; the variance of HR group is significantly less than Ctrl group for censoring distance values 2.4 *mm* and higher; and the variance of MDD group is greater than Ctrl group for $\gamma_{\delta,k} > 2.6$ *mm*. The interpretation of these variance differences is similar (i.e., larger variance implies more morphometric variability). ∎

*Remark 6:* **Handling Twin Dependence:** For the BF HOV test, we are actually performing ANOVA *F*-test on the residuals from the median, which is assuming within and between sample independence. In our example data set, the diagnostic groups consist of twin pairs. In particular, each MDD subject has a cotwin (i.e., the other member of the twin pair) who is labeled as HR (for MDD), and also, the Ctrl group consists of 14 twin pairs. Thus, we do not only have spatial dependence (due to the neighboring voxels) in the LCDM distances, but also (genetic) dependence between the LCDM distances of the cotwins. First observe that paired sample analysis cannot be performed on the distances of cotwins, since the number of LCDM distances do not match for the cotwins, so one cannot take differences of the distances and then analyze. We can view the distance measurements on cotwins as repeated measures, but the distance data is not balanced (that is, the number of distances (i.e., replications) for each subject is different). However, one can still account for such dependence by using a linear mixed effects model. In particular, in the "lme4" package in R, one can use "lmer" command with properly declaring the error structure. For example, let "resid" be residuals from the median, "lab" stand for the diagnostic labels, and "twin" for the twin factor (i.e., takes the same value for each twin pair), then the usual BF HOV test (assuming independence) would be based on the linear modeling of residuals with the labels as "anova(lm(resid ~ lab))". However, to take the twin dependence into account, we can perform mixed modeling as "lmer(resid ~ lab + (1|twin))". In particular, without accounting for twin dependence, BF HOV test yields the test statistic *F*=216.98, $p < .0001$, and with the mixed models approach, we obtain *t*=-5.66, $p < .0001$ (which is for comparing MDD residuals with HR residuals). This comparison suggests that although BF HOV test and the distribution/location tests are robust to non-normality and mild within sample dependence (due to spatial dependence from nearby voxels), between sample dependence (as in MDD vs HR) and within sample dependence as in the dependence between cotwins among Ctrl subjects seem to influence these tests much more severely. However, our goal in this article is not to tackle twin or other types of (between) sample dependence structures, but assess the application of the BF HOV test on the pooled and censored distances. So in our illustration of the method on the real life data, we ignore the twin dependence, as the methodology is intended to be applied to data that satisfies between sample independence. Yet, for interested readers, we also point out how to address such dependence in this remark. ∎

## 5. Discussion and Conclusions

In this article, we use LCDM distances to detect morphometric variability in brain tissues related to diseases. The ROI is the GM tissue in Ventral Medial Prefrontal Cortices (VMPFCs) for three groups of subjects; namely, subjects with major depressive disorder (MDD), subjects with high risk (HR) of MDD, and healthy control subjects (Ctrl). Our study comprises of (MDD, HR) and (Ctrl,Ctrl) cotwin pairs. Previously, to extract more information from the LCDM distances, pooling of the LCDM distances by group was recommended [8]. The pooled distances have been shown to be a powerful method to detect group differences in morphometry and stochastic ordering of the distances. In the same reference, it has also been shown that pooled LCDM distances are not significantly affected by the assumption violations such as within sample dependence or non-Gaussianity of the distances. Moreover, to determine at which distance from GM/WM surface the significant differences occur, censoring of LCDM distances was proposed [20], where it was also shown that the effect of assumption violations and data aggregation (due to censoring) has negligible effect on the inference based on censored LCDM distances.

For the current analyses, to test the homogeneity of the variances (HOV), we employ Brown-Forsythe (BF) test. BF HOV test is equivalent to applying ANOVA on the absolute difference of each distance (i.e., residuals) from the median. Hence BF test requires within sample independence and Gaussianity not for the individual distances, but for the residuals. However, if the raw distances satisfy these assumptions, so would the residuals. For LCDM data, within sample independence is violated due to the spatial correlation between LCDM distances that are from close-by voxels and Gaussianity is violated due to severe right skew of the distances. However, our Monte Carlo study shows that the influence of these violations is almost negligible for the BF test (as is the case for other tests (see [8]).

We demonstrate that HOV analysis of the pooled and censored LCDM distances yields important complementary information to the other tests of location (like ANOVA *F*-tests, or *t*-tests) or tests of distribution (like KW test, WRS test, and KS test) and is a powerful tool to detect morphometric variability. That is, HOV methodology is suggested as a complementary tool to the analyses as done in [8] and [20], and it gives information on "morphometric variability"; say for two groups A and B, if group A has distances with less variation than group B, then the morphometry of group A is closer to each other with more common defining characteristics. Furthermore, morphometric inference for group A would be more precise, which would be important in potential research or clinical applications. The HOV analysis on LCDM distances indicate that variability of left and right distances tend to decrease due to MDD or HR and the morphometric variation is smallest for the HR subjects for both left and right VMPFCs possibly due to thinning in left and right VMPFCs. Then the morphometry of MDD and HR subjects is less variable compared to healthy subjects, which indicates that statistical inference based on LCDM distances will be more reliable for these groups.

Using pooled LCDM distances, we can obtain an overall assessment of the morphometric variability, but not of the location of such differences, which could be crucial for understanding the underlying neurobiology. Hence, we also use censored LCDM distances for HOV analysis. That is, we also perform tests of equality of variances (i.e., HOV) of censored LCDM distances to detect the location (i.e., distance with respect to the GM/WM surface) of morphometric variations in cortex for GM start to be significant due to various conditions or associated with specific diseases. When the pooled LCDM distances are censored, we only keep distances up to censoring distance values $\gamma_{\delta,k} \in \{0.01, 0.02,...,5.50\}$. The censored distances (i.e., distances in $[0.5, \gamma_{\delta,k}]$ mm) can be used to determine the distance values (from GM/WM surface) at which significant group differences in variation can occur. In addition to the assumption violations for the pooled distances, there is also the issue of data aggregation at each censoring step. Our extensive Monte Carlo simulation study suggests that the influences of the assumption violations and data aggregation on HOV analysis of censored distances are negligible. We observe that BF HOV test and the other tests considered have significantly different rejection and acceptance regions, hence do not provide the same information. So we recommend BF tests in addition to other tests (of location or distribution) for censored LCDM distances, since they can help better understand the effect of the disease on the tissue in question.

The HOV analysis of the censored distances can be implemented in a pointwise fashion. Although our extensive Monte Carlo simulation study indicates that it also captures the distance ranges of significant differences (i.e., distances at which HOV is significantly violated) correctly. To have more reliable simultaneous inference, we discuss and apply various correction procedures for multiple testing. Among the correction methods we demonstrate that Benjamini-Hochberg (BH) procedure seems to be the most appropriate, as it maintains the correct conclusions for the null and alternative cases. The other methods are either extremely conservative (e.g., Bonferroni or Holm) or require independence (e.g., Šidák and Hochberg) of the tests performed. BH method also requires independence but is also valid for a special form of dependence, and there is evidence that censored LCDM distances satisfy this type of dependence (called positive regression dependency in [52]).

We also apply the methodology on simulated data sets from normal and exponential distributions. The pooled normal data satisfy all the assumptions (i.e., normality, within and between sample independence) and the pooled exponential data satisfy all assumptions except normality. On the other hand censored data violates normality for all censoring steps for non-normal distributions and except the last few steps for normal distributions. The pooled analysis on these data sets yield expected overall results (i.e., reflects the overall differences by design). However, the censoring analysis should be carefully interpreted and its local nature should always be taken into account. We observe that when more of the assumptions hold parametric tests tend to have higher power in pooled analysis; however, we also observe that the alternative parameters (that influence scale and location of the data) are more decisive on the performance of the tests compared to the distribution of the data sets. For example, at one type of alternatives, the higher power is attained by the exponential data while at another type higher power is attained by the normal data. In choosing the lowest (i.e., the first) censoring threshold we recommend the following approach. If data supports are bounded and are known, pick the largest of the infimum of the supports as the first censoring threshold, if the data support is unbounded below (as in the normal case), pick the lowest value that guarantees about 5-10 values for each sample when this value is used as a threshold, or pick a value close to the largest $\mu - 2.3\sigma$ values for the samples in the normal case. In practice, when the data supports are unknown, one can also choose the lowest value that guarantees about 5-10 values are available for the data analyses/comparisons. Also for some data there is a natural choice like the value 0 as in the LCDM data set.

Finally, we emphasize that the methodology used in this article for analyzing VMPFC shape variation differences is valid for application in other structures that lend themselves for measuring LCDM distances.

## Conflict of Interest Statement

The authors declare that there is no conflict of interests regarding the publication of this paper.

## Acknowledgements

Research supported by R01-MH62626-01, P41-EB015909, R01-MH57180 and EC was supported by a Marie Curie International Outgoing Fellowship within the 7[th] European Community Framework Programme (329370 PRinHDD).

## List of Abbreviations

LCDM: Labeled Cortical Distance Map
GM: Gray Matter; WM: White Matter; CSF: Cerebrospinal Fluid
BF Test: Brown-Forsythe Test
HOV: Homogeneity of Variance
VMPFC: Ventral Medial Prefrontal Cortex
MDD: Major Depressive Disorder; HR: High Risk; Ctrl: Control or Healthy
CA: Computational Anatomy
MRI: Magnetic Resonance Imaging
ROI: Region of Interest
KW Test: Kruskal-Wallis Test
WRS Test: Wilcoxon Rank Sum Test
KS Test: Kolmogorov-Smirnov Test
BH Procedure: Benjamini-Hochberg Procedure

BY Procedure: Benjamini-Yekutieli Procedure

# Tables

| $(n_x, n_y, n_z)$ | Empirical size | | | | Proportion of agreement | | | | | |
|---|---|---|---|---|---|---|---|---|---|---|
| | $\hat{\alpha}_{BF}$ | $\hat{\alpha}_{KW}$ | $\hat{\alpha}_{F_1}$ | $\hat{\alpha}_{F_2}$ | $\hat{\alpha}_{KW,F_1}$ | $\hat{\alpha}_{KW,F_2}$ | $\hat{\alpha}_{F_1,F_2}$ | $\hat{\alpha}_{BF,KW}$ | $\hat{\alpha}_{BF,F_1}$ | $\hat{\alpha}_{BF,F_2}$ |
| (1000,1000,1000) | .0495$^a$ | .0512$^a$ | .0501$^a$ | .0506$^a$ | .0403$^\ell$ | .0406$^\ell$ | .0492$^\approx$ | .0049$^\ell$ | .0080$^\ell$ | .0080$^\ell$ |
| (5000,5000,10000) | .0509$^a$ | .0474$^a$ | .0473$^a$ | .0477$^a$ | .0376$^\ell$ | .0378$^\ell$ | .0468$^\approx$ | .0044$^\ell$ | .0067$^\ell$ | .0068$^\ell$ |
| (5000,7500,10000) | .0486$^a$ | .0506$^a$ | .0484$^{a,<}$ | .0484$^{a,<}$ | .0396$^\ell$ | .0395$^\ell$ | .0479$^\approx$ | .0055$^\ell$ | .0078$^\ell$ | .0078$^\ell$ |
| (10000,10000,10000) | .0512$^a$ | .0515$^a$ | .0501$^a$ | .0500$^a$ | .0404$^\ell$ | .0403$^\ell$ | .0497$^\approx$ | .0041$^\ell$ | .0074$^\ell$ | .0076$^\ell$ |

**Table 1:** Estimated significance levels (i.e., empirical size estimates) and proportions of agreement between the test pairs based on Monte Carlo simulations of distances with three groups, $\mathcal{X}$, $\mathcal{Y}$, and $\mathcal{Z}$

each of size $n_x$, $n_y$, and $n_z$, respectively, with $N_{mc}=10000$ Monte Carlo replicates. The empirical sizes in the same row with the same superscript are not significantly different from each other. Empirical size estimates within $[.0464,.0536]$ are not significantly different from the nominal level of 0.05. (In the superscripts ">" means the empirical size is significantly larger than 0.05; i.e., method is liberal; "<" means empirical size is significantly smaller than 0.05; i.e., method is conservative; "$\ell$" means the proportion of agreement significantly less than the minimum of the empirical sizes; "$\approx$" means the proportion of agreement not significantly less than the minimum of the corresponding empirical sizes.)

| | Two-sided Tests ||||||||||
|---|---|---|---|---|---|---|---|---|---|---|
| | Empirical size |||| Proportion of agreement ||||||
| $(n_x,n_y)$ | $\hat{\alpha}_{BF}$ | $\hat{\alpha}_W$ | $\hat{\alpha}_t$ | $\hat{\alpha}_{KS}$ | $\hat{\alpha}_{W,t}$ | $\hat{\alpha}_{W,KS}$ | $\hat{\alpha}_{t,KS}$ | $\hat{\alpha}_{BF,W}$ | $\hat{\alpha}_{BF,t}$ | $\hat{\alpha}_{BF,KS}$ |
| (1000,1000) | .0522$^a$ | .0482$^a$ | .0488$^a$ | .0471$^a$ | .0392$^\ell$ | .0301$^\ell$ | .0274$^\ell$ | .0055$^\ell$ | .0082$^\ell$ | .0091$^\ell$ |
| (5000,10000) | .0512$^a$ | .0462$^{a,<}$ | .0474$^a$ | .0460$^{a,<}$ | .0378$^\ell$ | .0276$^\ell$ | .0239$^\ell$ | .0050$^\ell$ | .0084$^\ell$ | .0079$^\ell$ |
| (7500,10000) | .0479$^a$ | .0511$^a$ | .0516$^a$ | .0479$^a$ | .0414$^\ell$ | .0315$^\ell$ | .0286$^\ell$ | .0048$^\ell$ | .0079$^\ell$ | .0088$^\ell$ |
| (10000,10000) | .0496$^a$ | .0476$^a$ | .0487$^a$ | .0447$^{a,<}$ | .0390$^\ell$ | .0267$^\ell$ | .0241$^\ell$ | .0046$^\ell$ | .0086$^\ell$ | .0079$^\ell$ |
| | Left-Sided Tests (i.e., values in $\mathcal{Y}$ tend to be smaller than those in $\mathcal{Z}$) ||||||||||
| | Empirical size |||| Proportion of agreement ||||||
| (1000,1000) | .0518$^a$ | .0435$^{b,<}$ | .0453$^{b,<}$ | .0432$^{b,<}$ | .0363$^\ell$ | .0278$^\ell$ | .0256$^\ell$ | .0044$^\ell$ | .0071$^\ell$ | .0071$^\ell$ |
| (5000,10000) | .0522$^a$ | .0481$^a$ | .0495$^a$ | .0485$^a$ | .0389$^\ell$ | .0300$^\ell$ | .0277$^\ell$ | .0035$^\ell$ | .0060$^\ell$ | .0063$^\ell$ |
| (7500,10000) | .0525$^a$ | .0510$^a$ | .0499$^a$ | .0478$^a$ | .0421$^\ell$ | .0326$^\ell$ | .0302$^\ell$ | .0044$^\ell$ | .0065$^\ell$ | .0067$^\ell$ |
| (10000,10000) | .0482$^a$ | .0473$^a$ | .0471$^a$ | .0460$^{a,<}$ | .0399$^\ell$ | .0297$^\ell$ | .0278$^\ell$ | .0048$^\ell$ | .0074$^\ell$ | .0081$^\ell$ |
| | Right-Sided Tests (i.e., values in values in $\mathcal{Y}$ tend to be larger than those in $\mathcal{Z}$) ||||||||||
| | Empirical size |||| Proportion of agreement ||||||
| (1000,1000) | .0507$^a$ | .0537$^a$ | .0539$^a$ | .0508$^a$ | .0436$^\ell$ | .0337$^\ell$ | .0311$^\ell$ | .0048$^\ell$ | .0079$^\ell$ | .0083$^\ell$ |
| (5000,10000) | .0513$^a$ | .0487$^a$ | .0486$^a$ | .0497$^a$ | .0400$^\ell$ | .0324$^\ell$ | .0304$^\ell$ | .0055$^\ell$ | .0083$^\ell$ | .0087$^\ell$ |
| (7500,10000) | .0461$^{a,<}$ | .0488$^a$ | .0501$^a$ | .0495$^a$ | .0401$^\ell$ | .0322$^\ell$ | .0305$^\ell$ | .0039$^\ell$ | .0062$^\ell$ | .0075$^\ell$ |
| (10000,10000) | .0500$^a$ | .0477$^a$ | .0481$^a$ | .0470$^a$ | .0395$^\ell$ | .0304$^\ell$ | .0290$^\ell$ | .0040$^\ell$ | .0067$^\ell$ | .0063$^\ell$ |

**Table 2:** Estimated significance levels (i.e., empirical size estimates) under $H_o$ where $r_y = r_z = 1.0$ and $\eta_y = \eta_z = 0$ based on Monte Carlo simulation of LCDM-like distances with two groups $\mathcal{Y}$ and $\mathcal{Z}$ each with size $n_y$ and $n_z$, respectively, with $N_{mc}=10000$ Monte Carlo replicates. The size estimates in the same row are superscripted so that order of significance difference is as the alphabetical order of the superscripts, and size estimates with the same superscript are not significantly different from each other. Superscripts for the proportions of agreement are as in Table 1.

| | $L1:(r_y,r_z,\eta_y,\eta_z)=(1.1,1.0,0,0)$ |||| $L2:(r_y,r_z,\eta_y,\eta_z)=(1.1,1.2,0,0)$ ||||
|---|---|---|---|---|---|---|---|---|
| $(n_x,n_y,n_z)$ | $\hat{\beta}_{BF}$ | $\hat{\beta}_{KW}$ | $\hat{\beta}_{F_1}$ | $\hat{\beta}_{F_2}$ | $\hat{\beta}_{BF}$ | $\hat{\beta}_{KW}$ | $\hat{\beta}_{F_1}$ | $\hat{\beta}_{F_2}$ |
| (1000,1000,1000) | .0511$^b$ | .0778$^a$ | .0770$^a$ | .0768$^a$ | .0516$^c$ | .1396$^a$ | .1316$^{ab}$ | .1313$^b$ |
| (5000,5000,10000) | .0511$^c$ | .2281$^a$ | .2137$^b$ | .2114$^b$ | .0519$^c$ | .6725$^a$ | .6315$^b$ | .6317$^b$ |
| (5000,7500,10000) | .0499 | .2903$^a$ | .2698$^b$ | .2694$^b$ | .0513$^c$ | .6842$^a$ | .6396$^b$ | .6401$^b$ |
| (10000,10000,10000) | .0482$^c$ | .3900$^a$ | .3564$^b$ | .3559$^b$ | .0490$^c$ | .8410$^a$ | .8050$^b$ | .8050$^b$ |
| | $L3:(r_y,r_z,\eta_y,\eta_z)=(1.0,1.0,10,0)$ |||| $L4:(r_y,r_z,\eta_y,\eta_z)=(1.0,1.0,10,30)$ ||||
| (1000,1000,1000) | .0899$^a$ | .0574$^c$ | .0728$^b$ | .0721$^b$ | .2236$^a$ | .0963$^c$ | .1519$^b$ | .1512$^b$ |
| (5000,5000,10000) | .3408$^a$ | .0767$^c$ | .1930$^b$ | .1854$^b$ | .9255$^a$ | .3986$^c$ | .7436$^b$ | .7537$^b$ |
| (5000,7500,10000) | .4378$^a$ | .0832$^c$ | .2415$^b$ | .2360$^b$ | .9214$^a$ | .4191$^c$ | .7578$^b$ | .7627$^b$ |
| (10000,10000,10000) | .5564$^a$ | .1006$^c$ | .3127$^b$ | .3061$^b$ | .9851$^a$ | .5352$^c$ | .8842$^b$ | .8835$^b$ |

**Table 3:** The power estimates based on Monte Carlo simulation of distances under the alternative cases L1-L4 with three groups, $\mathcal{X}$, $\mathcal{Y}$, and $\mathcal{Z}$ each with size $n_x$, $n_y$, and $n_z$, respectively, with $N_{mc}=10000$ Monte Carlo replicates. The power estimates in the same row are superscripted so that order of significance

difference is as the alphabetical order of the superscripts, and power estimates with the same superscript are not significantly different from each other.

| | L2: $(r_y, r_z, \eta_y, \eta_z) = (1.1, 1.2, 0, 0)$ | | | | L4: $(r_y, r_z, \eta_y, \eta_z) = (1.0, 1.0, 10, 30)$ | | | |
|---|---|---|---|---|---|---|---|---|
| | ($\mathcal{X} < \mathcal{Y}$) | | | | ($\mathcal{X} < \mathcal{Y}$) | | | |
| $(n_x, n_y, n_z)$ | $\hat{\beta}_{BF}$ | $\hat{\beta}_W$ | $\hat{\beta}_t$ | $\hat{\beta}_{KS}$ | $\hat{\beta}_{BF}$ | $\hat{\beta}_W$ | $\hat{\beta}_t$ | $\hat{\beta}_{KS}$ |
| (1000,1000,1000) | .0496[b] | .1231[a] | .1187[a] | .1198[a] | .1543[b] | .0714[c] | .1129[a] | .0601[d] |
| (5000,5000,10000) | .0591[d] | .3182[b] | .2975[c] | .3907[a] | .4201[a] | .1263[c] | .2698[b] | .0971[d] |
| (5000,7500,10000) | .0531[d] | .3445[b] | .3231[c] | .4514[a] | .4731[a] | .1302[c] | .2940[b] | .1019[d] |
| (10000,10000,10000) | .0528[d] | .4930[b] | .4618[c] | .6904[a] | .6494[a] | .1687[c] | .4137[b] | .1250[d] |
| | ($\mathcal{X} < \mathcal{Z}$) | | | | ($\mathcal{X} < \mathcal{Z}$) | | | |
| (1000,1000,1000) | .0531[c] | .2709[b] | .2602[b] | .2868[a] | .4029[a] | .1798[c] | .2945[b] | .1438[d] |
| (5000,5000,10000) | .0665[d] | .8422[b] | .8183[c] | .9753[a] | .9793[a] | .5770[c] | .8869[b] | .8874[b] |
| (5000,7500,10000) | .0633[d] | .8415[b] | .8181[c] | .9720[a] | .9788[a] | .5856[c] | .8895[b] | .8863[b] |
| (10000,10000,10000) | .0659[d] | .9428[b] | .9277[c] | .9987[a] | .9975[a] | .7349[d] | .9670[c] | .9947[b] |
| | ($\mathcal{Y} < \mathcal{Z}$) | | | | ($\mathcal{Y} < \mathcal{Z}$) | | | |
| (1000,1000,1000) | .0536[d] | .1368[a] | .1266[b] | .1323[ab] | .1791[a] | .1278[c] | .1562[b] | .1064[d] |
| (5000,5000,10000) | .0587[d] | .3950[b] | .3498[c] | .5084[a] | .6172[a] | .3513[d] | .4914[b] | .3909[c] |
| (5000,7500,10000) | .0577[d] | .4785[b] | .4269[c] | .6198[a] | .7098[a] | .4290[d] | .5923[c] | .5511[b] |
| (10000,10000,10000) | .0648[d] | .5338[b] | .4709[c] | .6896[a] | .7601[a] | .4740[c] | .6537[b] | .6484[b] |

**Table 4:** The power estimates of the tests for the pairwise comparisons of the samples based on Monte Carlo simulation of distances under the alternative cases *L2* and *L4* with three groups, $\mathcal{X}$, $\mathcal{Y}$, and $\mathcal{Z}$ each with size $n_x$, $n_y$, and $n_z$, respectively, with $N_{mc} = 10000$ Monte Carlo replicates. The superscripts are as in Table 3.

| | Normal Data | | | | Exponential Data | | | |
|---|---|---|---|---|---|---|---|---|
| $(n_x, n_y, n_z)$ | $\hat{\alpha}_{BF}$ | $\hat{\alpha}_{KW}$ | $\hat{\alpha}_{F_1}$ | $\hat{\alpha}_{F_2}$ | $\hat{\alpha}_{BF}$ | $\hat{\alpha}_{KW}$ | $\hat{\alpha}_{F_1}$ | $\hat{\alpha}_{F_2}$ |
| (1000,1000,1000) | .0498[a] | .0499[a] | .0503[a] | .0497[a] | .0455[b,<] | .0519[a] | .0492[ab] | .0498[ab] |
| (5000,5000,10000) | .0486[a] | .0529[a] | .0527[a] | .0522[a] | .0544[a] | .0500[a] | .0517[a] | .0524[a] |
| (5000,7500,10000) | .0530[a] | .0498[a] | .0519[a] | .0515[a] | .0481[a] | .0475[a] | .0463[a,<] | .0465[a] |
| (10000,10000,10000) | .0506[a] | .0511[a] | .0492[a] | .0492[a] | .0522[a] | .0531[a] | .0553[a,>] | .0550[a,>] |

**Table 5:** Estimated significance levels (i.e., empirical size estimates) based on Monte Carlo simulations of normal data (left) and exponential data (right) with three groups, $\mathcal{X}$, $\mathcal{Y}$, and $\mathcal{Z}$ each with size $n_x$, $n_y$, and $n_z$, respectively, with $N_{mc} = 10000$ Monte Carlo replicates. Superscript labeling is as in Table 1.

| | N1 | | | | E1 | | | |
|---|---|---|---|---|---|---|---|---|
| $(n_x, n_y, n_z)$ | $\hat{\beta}_{BF}$ | $\hat{\beta}_{KW}$ | $\hat{\beta}_{F_1}$ | $\hat{\beta}_{F_2}$ | $\hat{\beta}_{BF}$ | $\hat{\beta}_{KW}$ | $\hat{\beta}_{F_1}$ | $\hat{\beta}_{F_2}$ |
| (1000,1000,1000) | .0498[b] | .0660[a] | .0645[a] | .0652[a] | .0531[c] | .1410[a] | .0842[b] | .0857[b] |
| (5000,5000,10000) | .0486[b] | .1411[a] | .1453[a] | .1449[a] | .0556[c] | .6168[a] | .2811[b] | .2794[b] |
| (5000,7500,10000) | .0530[b] | .1760[a] | .1827[a] | .1829[a] | .0528[c] | .7312[a] | .3560[b] | .3574[b] |
| (10000,10000,10000) | .0506[b] | .2216[a] | .2283[a] | .2277[a] | .0549[c] | .8617[a] | .4561[b] | .4585[b] |
| | N2 | | | | E2 | | | |
| (1000,1000,1000) | .0498[b] | .0712[a] | .0688[a] | .0682[a] | .0555[c] | .3456[a] | .1620[b] | .1624[b] |
| (5000,5000,10000) | .0486[b] | .1836[a] | .1931[a] | .1934[a] | .0707[c] | .9916[a] | .7871[b] | .7877[b] |
| (5000,7500,10000) | .0530[b] | .1913[a] | .1940[a] | .1934[a] | .0631[c] | .9922[a] | .7940[b] | .7942[b] |
| (10000,10000,10000) | .0506[c] | .2864[b] | .2996[a] | .3000[a] | .0738[c] | .9997[a] | .9252[b] | .9252[b] |
| | N3 | | | | E3 | | | |

| $(n_x,n_y,n_z)$ | $\hat{\beta}_{BF}$ | $\hat{\beta}_W$ | $\hat{\beta}_t$ | $\hat{\beta}_{KS}$ | $\hat{\beta}_{BF}$ | $\hat{\beta}_W$ | $\hat{\beta}_t$ | $\hat{\beta}_{KS}$ |
|---|---|---|---|---|---|---|---|---|
| (1000,1000,1000) | .0948[a] | .0650[b] | .0642[b] | .0648[b] | .0881[a] | .0513[c] | .0722[b] | .0722[b] |
| (5000,5000,10000) | .3342[a] | .1396[b] | .1462[b] | .1423[b] | .2873[a] | .0657[c] | .1997[b] | .1906[b] |
| (5000,7500,10000) | .4269[a] | .1741[b] | .1802[b] | .1784[b] | .3515[a] | .0684[c] | .2423[b] | .2357[b] |
| (10000,10000,10000) | .5619[a] | .2178[b] | .2258[b] | .2228[b] | .4600[a] | .0742[c] | .3159[b] | .3070[b] |
| | N4 | | | | E4 | | | |
| (1000,1000,1000) | .1610[a] | .0863[b] | .0860[b] | .0861[b] | .1310[a] | .0696[c] | .1100[b] | .1115[b] |
| (5000,5000,10000) | .7543[a] | .3133[c] | .3257[bc] | .3336[b] | .6317[a] | .2194[c] | .5135[b] | .5270[b] |
| (5000,7500,10000) | .7618[a] | .3153[c] | .3231[bc] | .3313[b] | .6378[a] | .2340[c] | .5172[b] | .5311[b] |
| (10000,10000,10000) | .9052[a] | .4546[c] | .4719[b] | .4741[b] | .8214[a] | .3078[c] | .6968[b] | .7010[b] |

**Table 6:** The power estimates for the multi-group tests based on Monte Carlo simulation of normal data under the given alternatives $N1-N4$ (left) and exponential data under the alternatives $E1-E4$ (right) with samples $\mathcal{X}$, $\mathcal{Y}$, and $\mathcal{Z}$ each with size $n_x$, $n_y$, and $n_z$, respectively, with $N_{mc}=10000$ Monte Carlo replicates. The superscripts are as in Table 3.

| | Normal Data | | | | | | | |
|---|---|---|---|---|---|---|---|---|
| | $N2: \mu_x=3.35, \mu_y=3.39, \mu_z=3.40;$ $\sigma_x=\sigma_y=\sigma_z=2.28$ | | | | $N4: \mu_x=3.35, \mu_y=3.39, \mu_z=3.42;$ $\sigma_x=2.28, \sigma_y=2.33, \sigma_z=2.37$ | | | |
| | $\mathcal{X}<\mathcal{Y}$ | | | | $\mathcal{X}<\mathcal{Y}$ | | | |
| $(n_x,n_y,n_z)$ | $\hat{\beta}_{BF}$ | $\hat{\beta}_W$ | $\hat{\beta}_t$ | $\hat{\beta}_{KS}$ | $\hat{\beta}_{BF}$ | $\hat{\beta}_W$ | $\hat{\beta}_t$ | $\hat{\beta}_{KS}$ |
| (1000,1000,1000) | .0501[b] | .1008[a] | .1070[a] | .0942[a] | .1535[a] | .1001[b] | .1064[b] | .0997[b] |
| (5000,5000,10000) | .0519[c] | .2081[a] | .2144[a] | .1772[b] | .4160[a] | .2054[c] | .2128[b] | .2198[b] |
| (5000,7500,10000) | .0497[c] | .2435[a] | .2484[a] | .2062[b] | .4694[a] | .2400[c] | .2468[c] | .2642[b] |
| (10000,10000,10000) | .0516[c] | .3325[a] | .3455[a] | .2790[b] | .6550[a] | .3280[c] | .3397[c] | .3861[b] |
| | $\mathcal{X}<\mathcal{Z}$ | | | | $\mathcal{X}<\mathcal{Z}$ | | | |
| (1000,1000,1000) | .0523[c] | .1225[a] | .1236[a] | .1121[b] | .3149[a] | .1592[b] | .1650[b] | .1635[b] |
| (5000,5000,10000) | .0478[c] | .3430[a] | .3546[a] | .2867[b] | .9057[a] | .5258[d] | .5451[c] | .6696[b] |
| (5000,7500,10000) | .0482[c] | .3463[a] | .3562[a] | .2972[b] | .9074[a] | .5271[d] | .5480[c] | .6768[b] |
| (10000,10000,10000) | .0507[c] | .4547[a] | .4660[a] | .3790[b] | .9767[a] | .6722[d] | .6913[c] | .8385[b] |
| | $\mathcal{Y}<\mathcal{Z}$ | | | | $(\mathcal{Y}<\mathcal{Z}$ | | | |
| (1000,1000,1000) | .0529[b] | .0619[a] | .0636[a] | .0574[ab] | .1303[a] | .0866[bc] | .0910[b] | .0822[c] |
| (5000,5000,10000) | .0477[c] | .0869[a] | .0884[a] | .0766[b] | .3615[a] | .1847[bc] | .1930[b] | .1981[b] |
| (5000,7500,10000) | .0479[b] | .0869[a] | .0872[a] | .0825[a] | .4287[a] | .2070[c] | .2106[c] | .2255[b] |
| (10000,10000,10000) | .0458[c] | .0919[a] | .0935[a] | .0827[b] | .4688[a] | .2302[c] | .2323[c] | .2496[b] |

**Table 7:** The power estimates for pairwise tests based on Monte Carlo simulation of normal data under the given alternatives $N2$ and $N4$ with samples $\mathcal{X}$, $\mathcal{Y}$, and $\mathcal{Z}$ each with size $n_x$, $n_y$, and $n_z$, respectively, with $N_{mc}=10000$ Monte Carlo replicates. The superscripts are as in Table 3.

| | Exponential Data | | | | | | | |
|---|---|---|---|---|---|---|---|---|
| | $E2:(r_y,r_z,\eta_y,\eta_z)=(1.1,1.2,0,0)$ | | | | $E4:(r_y,r_z,\eta_y,\eta_z)=(1.0,1.0,10,30)$ | | | |
| | $\mathcal{X}<\mathcal{Y}$ | | | | $\mathcal{X}<\mathcal{Y}$ | | | |
| $(n_x,n_y,n_z)$ | $\hat{\beta}_{BF}$ | $\hat{\beta}_W$ | $\hat{\beta}_t$ | $\hat{\beta}_{KS}$ | $\hat{\beta}_{BF}$ | $\hat{\beta}_W$ | $\hat{\beta}_t$ | $\hat{\beta}_{KS}$ |
| (1000,1000,1000) | .0587[d] | .2346[b] | .1441[c] | .3246[a] | .1410[a] | .0618[c] | .1189[b] | .0541[d] |
| (5000,5000,10000) | .0773[d] | .6650[b] | .3748[c] | .9292[a] | .3651[a] | .0957[c] | .2806[b] | .0731[d] |
| (5000,7500,10000) | .0734[d] | .7197[b] | .4077[c] | .9643[a] | .4081[a] | .0981[c] | .3047[b] | .0751[d] |
| (10000,10000,10000) | .0813[d] | .8932[b] | .5834[c] | .9979[a] | .5714[a] | .1143[c] | .4338[b] | .0833[d] |
| | $\mathcal{X}<\mathcal{Z}$ | | | | $\mathcal{X}<\mathcal{Z}$ | | | |
| (1000,1000,1000) | .0689[d] | .5530[b] | .3206[c] | .8004[a] | .2618[a] | .1326[c] | .2194[b] | .1090[d] |
| (5000,5000,10000) | .1203[d] | .9983[b] | .9129[c] | 1.000[a] | .8352[a] | .3799[d] | .7356[b] | .5354[c] |
| (5000,7500,10000) | .1194[d] | .9984[b] | .9153[c] | 1.000[a] | .8399[a] | .3783[d] | .7388[b] | .5294[c] |
| (10000,10000,10000) | .1398[c] | .9999[a] | .9808[b] | 1.000[a] | .9413[a] | .4984[d] | .8689[b] | .7727[c] |
| | $\mathcal{Y} < \mathcal{Z}$ | | | | $\mathcal{Y} < \mathcal{Z}$ | | | |
| (1000,1000,1000) | .0619[d] | .2334[b] | .1449[c] | .3181[a] | .1081[a] | .1004[a] | .1072[a] | .0880[b] |
| (5000,5000,10000) | .0753[d] | .7689[b] | .4367[c] | .9697[a] | .2795[a] | .2318[c] | .2651[b] | .2547[b] |
| (5000,7500,10000) | .0864[d] | .8549[b] | .5329[c] | .9923[a] | .3437[a] | .2848[c] | .3239[b] | .3540[a] |
| (10000,10000,10000) | .0906[d] | .8987[b] | .5879[c] | .9971[a] | .3736[b] | .3122[d] | .3578[c] | .4207[a] |

**Table 8:** The power estimates based on Monte Carlo simulation of exponential data under the given alternatives $E2$ and $E4$ with samples $\mathcal{X}$, $\mathcal{Y}$, and $\mathcal{Z}$ each with size $n_x$, $n_y$, and $n_z$, respectively, with $N_{mc} = 10000$ Monte Carlo replicates. The superscripts are as in Table 3.

| | \multicolumn{6}{c}{Normal Data} | | | |
|---|---|---|---|---|---|---|
| | KW test | | ANOVA $F$-test | | BF HOV test | |
| case | range | $\widehat{\beta}_{\max}$ | range | $\widehat{\beta}_{\max}$ | range | $\widehat{\beta}_{\max}$ |
| $N1$ | (3.34,10.5) | .223 | (.18,10.5) | .236 | (.82,4.66) | .088 |
| $N2$ | (1.56,10.5) | .283 | (.12,10.5) | .284 | (.66,6.58)∪(7.86,9.04) | .087 |
| $N3$ | $(0,10.5) \setminus (5.04, 6.58)$ | .209 | $(0,10.5) \setminus (5.42, 6.96)$ | .215 | (0,10.5) | .518 |
| $N4$ | $(0,10.5) \setminus (5.34, 6.28)$ | .403 | $(0,10.5) \setminus (5.68, 6.84)$ | .400 | (0,10.5) | .886 |

**Table 9:** The ranges of data in which censoring analysis yields significantly higher power than nominal level of .05 together with the maximum power (denoted $\widehat{\beta}_{\max}$) attained in these intervals. We compare three groups from normal distribution with KW, ANOVA $F$, and BF HOV tests.

| | \multicolumn{6}{c}{Normal Data, $\mathcal{X}$ vs $\mathcal{Y}$} | | | | | |
|---|---|---|---|---|---|---|
| cases | WRS test | | Welch's t test | | BF HOV test | |
| | $\mathcal{X} < \mathcal{Y}$ | $\mathcal{X} > \mathcal{Y}$ | $\mathcal{X} < \mathcal{Y}$ | $\mathcal{X} > \mathcal{Y}$ | $\mathcal{X} < \mathcal{Y}$ | $\mathcal{X} > \mathcal{Y}$ |
| | range; $\widehat{\beta}_{\max}$ | range; $\widehat{\beta}_{\max}$ | range; $\widehat{\beta}_{\max}$ | range; $\widehat{\beta}_{\max}$ | range; $\widehat{\beta}_{\max}$ | range; $\widehat{\beta}_{\max}$ |
| N1,N2 | (0,10.5);.344 | ---;.049 | (0,10.5);.347 | ---;.048 | ---;.058 | (0,8.76);.114 |
| N3,N4 | (5.9,10.5);.324 | (0,5.36);.180 | (6.38,10.5);.320 | (0,5.86);.232 | (0,10.5);.613 | ---;.023 |
| | \multicolumn{6}{c}{Normal Data, $\mathcal{X}$ vs $\mathcal{Z}$} | | | | | |
| | $\mathcal{X} < \mathcal{Z}$ | $\mathcal{X} > \mathcal{Z}$ | $\mathcal{X} < \mathcal{Z}$ | $\mathcal{X} > \mathcal{Z}$ | $\mathcal{X} < \mathcal{Z}$ | $\mathcal{X} > \mathcal{Z}$ |
| N1,N3 | ---;.066* | ---;.062 | ---;.064* | ---;.062 | ---;.061 | ---;.064* |
| N2 | (0,10.5);.431 | ---;.047 | (0,10.5);.444 | ---;.040 | ---;.048 | (0,8.94);.137 |
| N4 | (5.94,10.5);.64 | (0,5.62);.39 | (6.54,10.5);.636 | (0,6.20);.512 | (0,10.5);.971 | ---;.006 |
| | \multicolumn{6}{c}{Normal Data, $\mathcal{Y}$ vs $\mathcal{Z}$} | | | | | |
| | $\mathcal{Y} < \mathcal{Z}$ | $\mathcal{Y} > \mathcal{Z}$ | $\mathcal{Y} < \mathcal{Z}$ | $\mathcal{Y} > \mathcal{Z}$ | $\mathcal{Y} < \mathcal{Z}$ | $\mathcal{Y} > \mathcal{Z}$ |
| N1 | ---;.060 | (0,10.5);.341 | ---;.053 | (0,10.5);.347 | (.34,8.52);.116 | ---;.060 |
| N2 | (3.80,10.5);.086 | ---;.062 | (0,10.5);.080 | ---;.062 | ---;.062 | **;.073 |
| N3 | (0,5.38);.206 | (5.9,10.5);.324 | (0,6.02);.255 | (6.7,10.5);.325 | ---;.017 | (0,10.5);.646 |
| N4 | (6.34,10.5);.195 | (0,5.7);.174 | (7.18,10.5);.194 | (0,6.30);.202 | (0,10.5);.438 | ---;.024 |

**Table 10:** The ranges of data for which censoring analysis yields significantly higher power than nominal level of .05 together with the maximum power $\widehat{\beta}_{\max}$ attained in these intervals. We perform pairwise comparisons for the three groups from normal distribution in Table 9 with WRS, $t$, and BF HOV tests. * On a few of the censoring values, the test has power slightly higher than .05.

| | Left VMPFC | | | | Right VMPFC | | | |
|---|---|---|---|---|---|---|---|---|
| Group | $n$ | Mean | median | SD | $n$ | mean | median | SD |
| MDD | 238937 | 1.62 | 1.46 | 1.13 | 170534 | 1.63 | 1.49 | 1.10 |
| HR | 228224 | 1.61 | 1.46 | 1.11 | 216978 | 1.59 | 1.46 | 1.08 |
| Ctrl | 308498 | 1.66 | 1.50 | 1.14 | 293479 | 1.66 | 1.53 | 1.12 |
| Overall | 775659 | 1.63 | 1.48 | 1.13 | 680991 | 1.63 | 1.50 | 1.10 |

**Table 11:** The sample sizes ($n$), means, medians, and standard deviations (SD) of the pooled LCDM distances (in *mm*) for left and right VMPFCs categorized by group.

| | Left VMPFC | | | | |
|---|---|---|---|---|---|
| Pair | BF HOV test | WRS test | $t$-test | KS test | |
| MDD, HR | <.0001* ($g$) | .3022 ($\ell$) | .0383* ($g$) | <.0001* ($\ell$) | .0073* ($g$) |
| MDD, Ctrl | <.0001* ($\ell$) | <.0001* ($\ell$) | <.0001* ($\ell$) | .5362 ($\ell$) | <.0001* ($g$) |
| HR, Ctrl | <.0001* ($\ell$) | <.0001* ($\ell$) | <.0001* ($\ell$) | .4170 ($\ell$) | <.0001* ($g$) |
| | Right VMPFC | | | | |
| MDD, HR | <.0001* ($g$) | .0776 ($g$) | .0041* ($g$) | .0158* ($\ell$) | .6017 ($g$) |
| MDD, Ctrl | <.0001* ($\ell$) | <.0001* ($\ell$) | <.0001* ($\ell$) | .0069* ($\ell$) | <.0001* ($g$) |
| HR, Ctrl | <.0001* ($\ell$) | <.0001* ($\ell$) | <.0001* ($\ell$) | .0043* ($\ell$) | <.0001* ($g$) |

**Table 12:** The $p$-values for the pairwise comparisons of the pooled distances by WRS test, $t$-test, and BF HOV test. The $p$-values are adjusted by Holm's correction method. ($g$ = first group is larger than the second and $\ell$ = first group is smaller than the second.) The significant $p$-values are marked with an asterisk.

# Figures

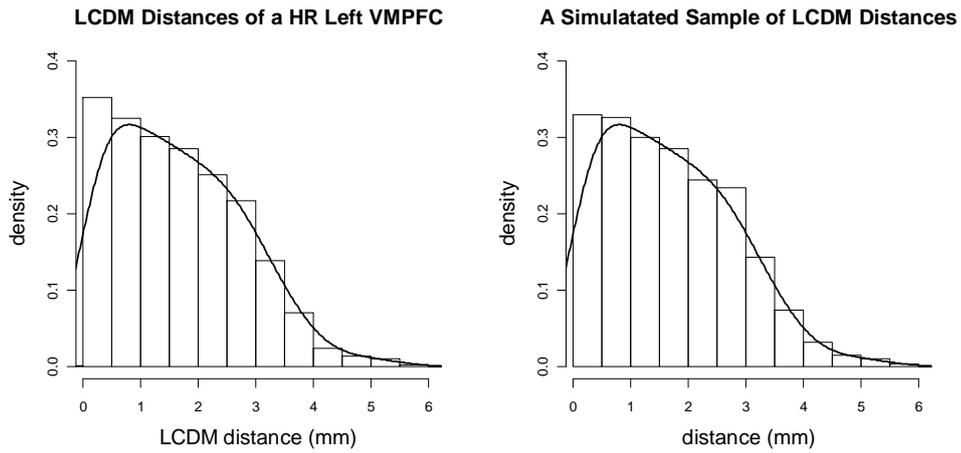

**Figure 1:** Histograms overlaid with the kernel density estimates of LCDM distances for the left VMPFC of HR subject 1 (left) and a simulated sample as described in Section 3.1.

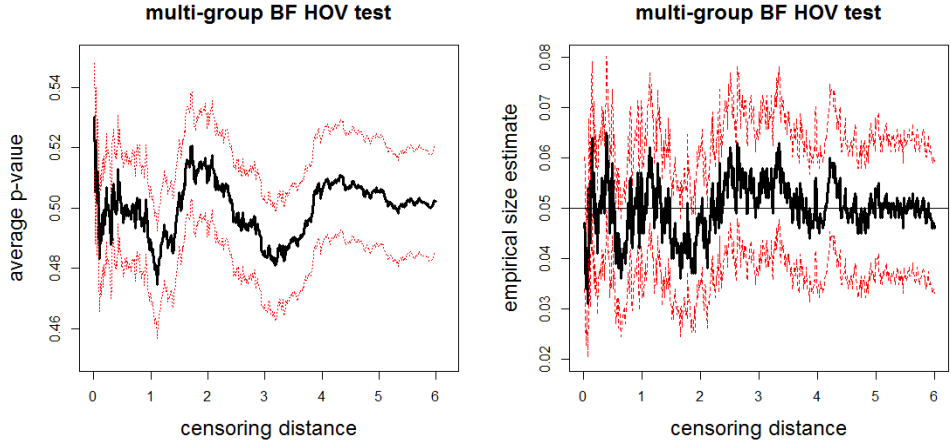

**Figure 2:** The average $p$-values (left) and empirical size estimates (right) together with 95% confidence bands versus censoring distance values under the null case for multi-group BF HOV test.

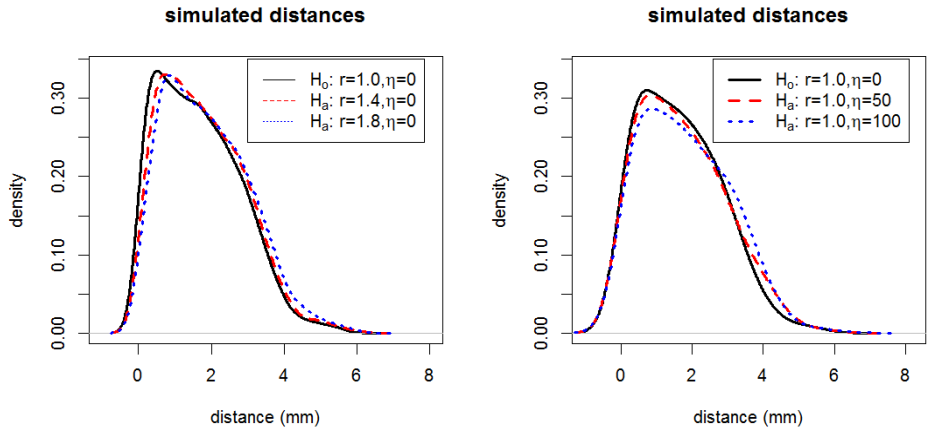

**Figure 3:** Plots of the kernel density estimates of the Monte Carlo simulated LCDM distances under the null case and alternatives with $\eta = 0$ and $r \in \{1.0, 1.4, 1.8\}$ (left); null case and alternatives with $r = 1.0$ and $\eta \in \{0, 50, 100\}$ (right). For the parameters $r$ and $\eta$, see Section 3.3.1.

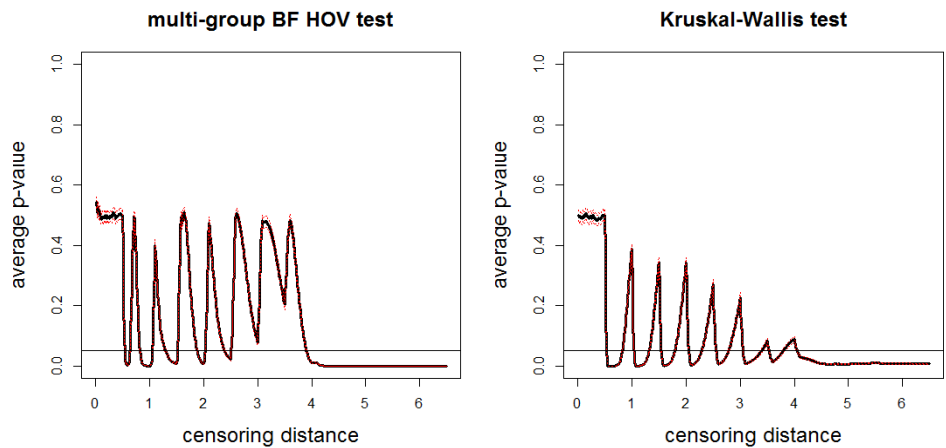

**Figure 4:** The average $p$-values versus censoring distances for multi-group BF HOV test (left) and multi-group KW test (right) together with the 95% confidence bands (dashed lines) based on 10000 Monte Carlo replications of censored $\mathcal{X}$, $\mathcal{Y}$, and $\mathcal{Z}$ sets that are generated under the alternative case L5. Horizontal lines are at 0.05.

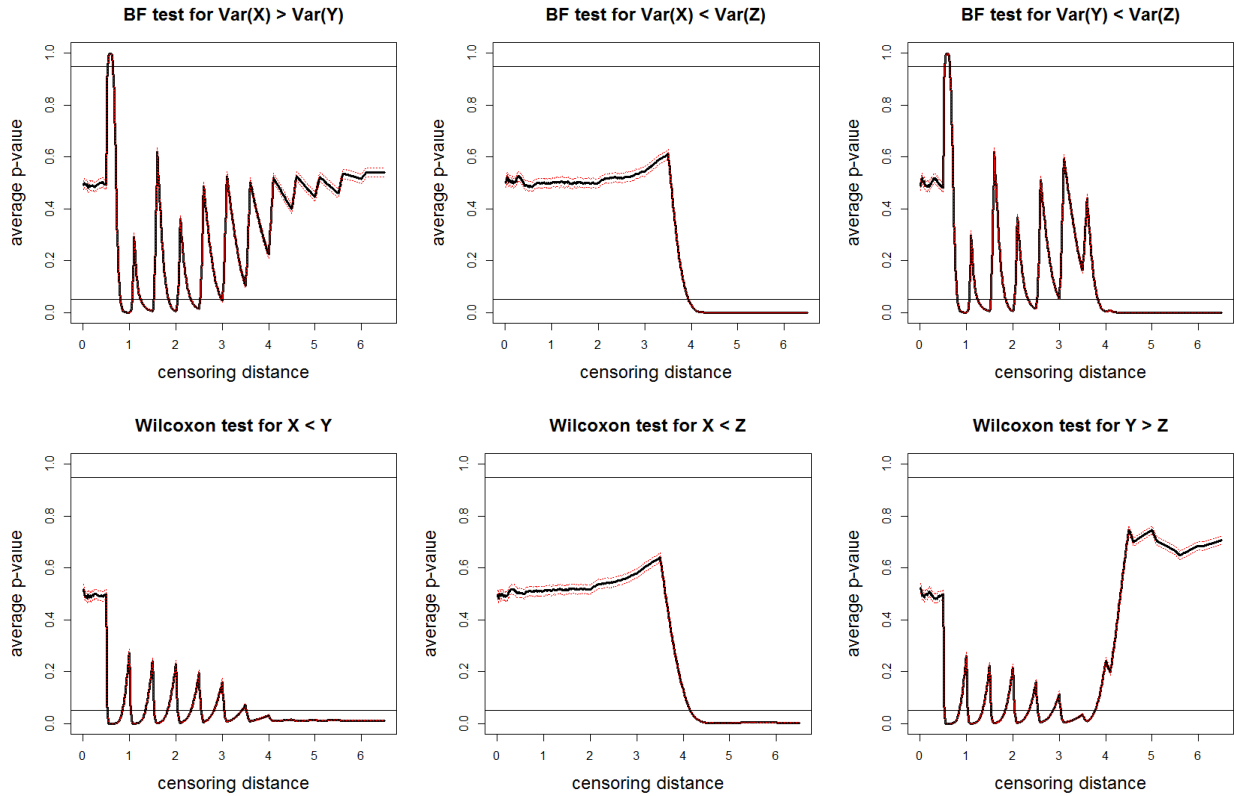

**Figure 5:** The average $p$-values together with 95% confidence bands versus censoring distance values under the alternative case L5 based on 10000 Monte Carlo replications for pairwise BF HOV test for the one-sided alternatives $\text{Var}(X) > \text{Var}(Y)$, $\text{Var}(X) < \text{Var}(Z)$, and $\text{Var}(Y) < \text{Var}(Z)$ and for pairwise WRS tests for the one-sided alternatives $X < Y$, $X < Z$, and $Y > Z$.

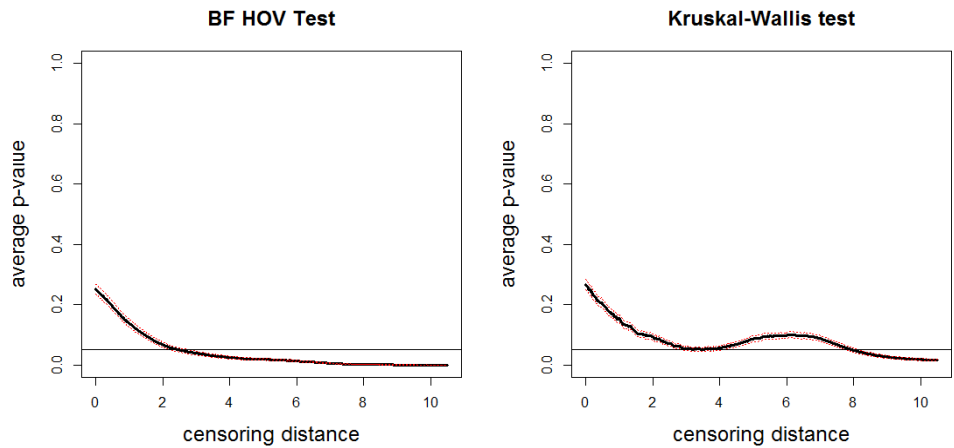

**Figure 6:** The average $p$-values versus censoring distances for multi-group BF HOV test (left) and multi-group KW test (right) together with the 95% confidence bands (dashed lines) based on 10000 Monte Carlo replications of censored $\mathcal{X}$, $\mathcal{Y}$, and $\mathcal{Z}$ sets that are generated under the alternative case N5 for normal data. Horizontal lines are at 0.05.

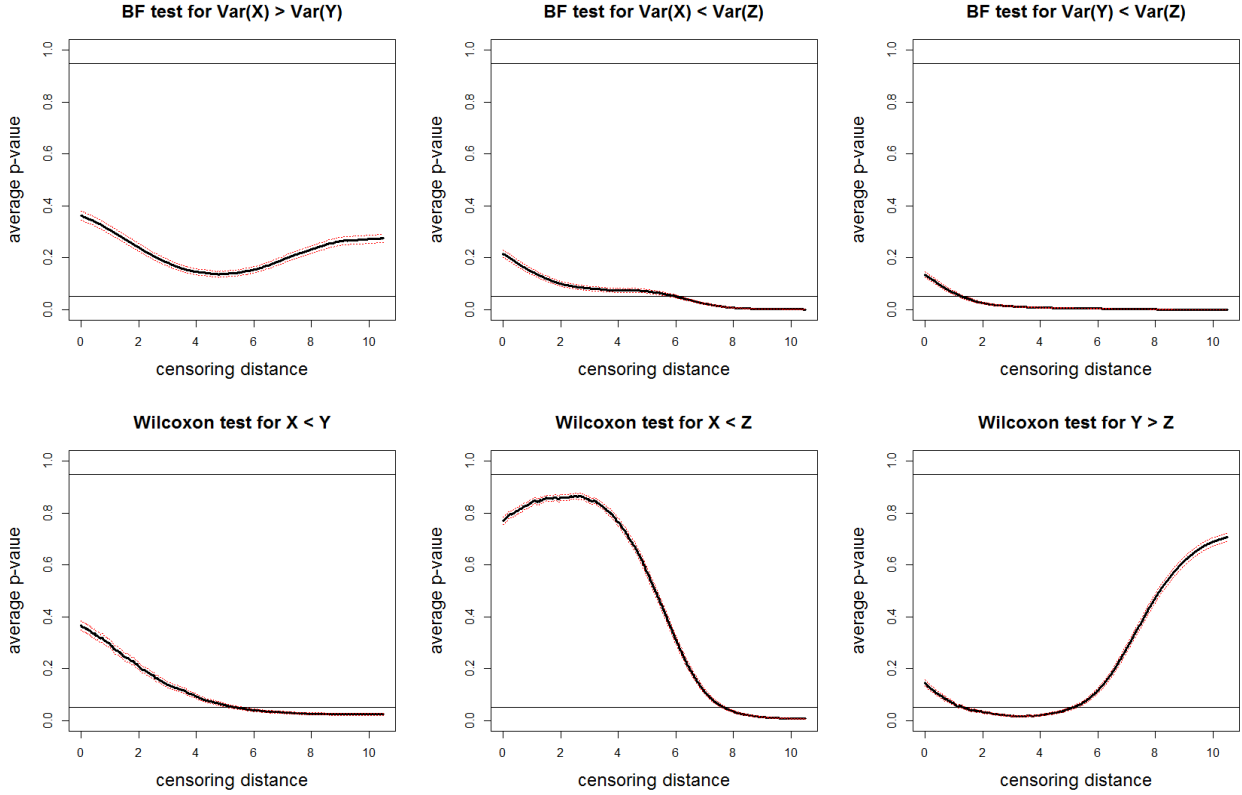

**Figure 7:** The average $p$-values together with 95% confidence bands versus censoring distance values under the alternative case N5 with normal data based on 10000 Monte Carlo replications for pairwise BF HOV test for the one-sided alternatives $\mathrm{Var}(X) > \mathrm{Var}(Y)$, $\mathrm{Var}(X) < \mathrm{Var}(Z)$, and $\mathrm{Var}(Y) < \mathrm{Var}(Z)$ and for pairwise WRS tests for the one-sided alternatives $X < Y$, $X < Z$, and $Y > Z$.

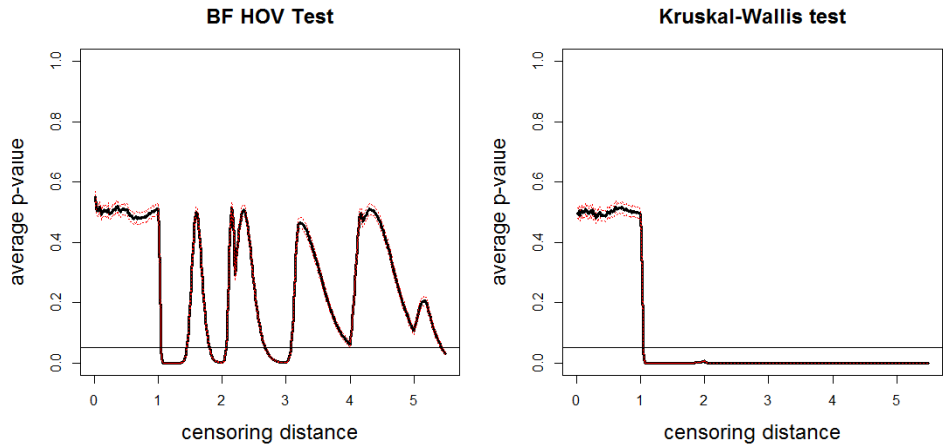

**Figure 8:** The average $p$-values versus censoring distances for multi-group BF HOV test (left) and multi-group KW test (right) together with the 95% confidence bands (dashed lines) based on 10000 Monte Carlo replications of censored $\mathcal{X}$, $\mathcal{Y}$, and $\mathcal{Z}$ sets that are generated under the alternative case E5 for exponential data. Horizontal lines are at 0.05.

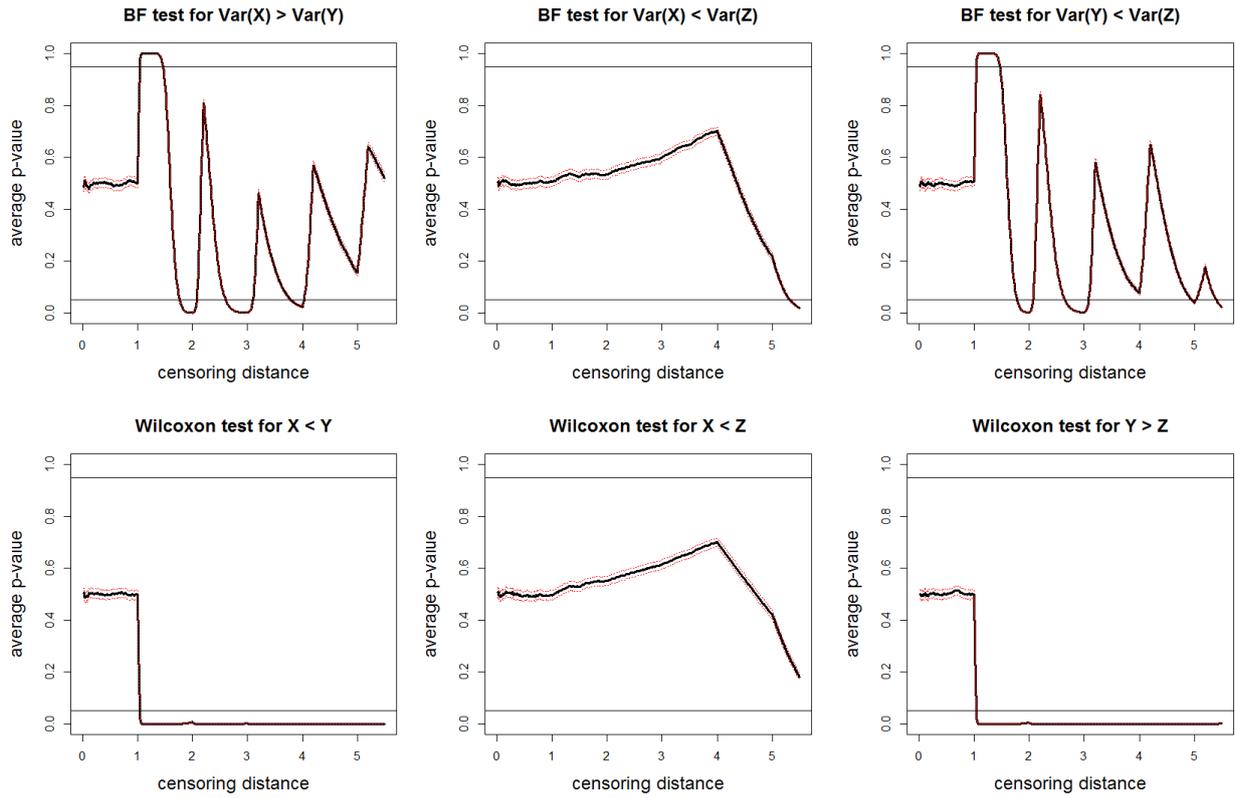

**Figure 9:** The average $p$-values together with 95% confidence bands versus censoring distance values under the alternative case E5 with exponential data based on 10000 Monte Carlo replications for pairwise BF HOV test for the one-sided alternatives $\mathrm{Var}(X) > \mathrm{Var}(Y)$, $\mathrm{Var}(X) < \mathrm{Var}(Z)$, and $\mathrm{Var}(Y) < \mathrm{Var}(Z)$ and for pairwise WRS tests for the one-sided alternatives $X < Y$, $X < Z$, and $Y > Z$.

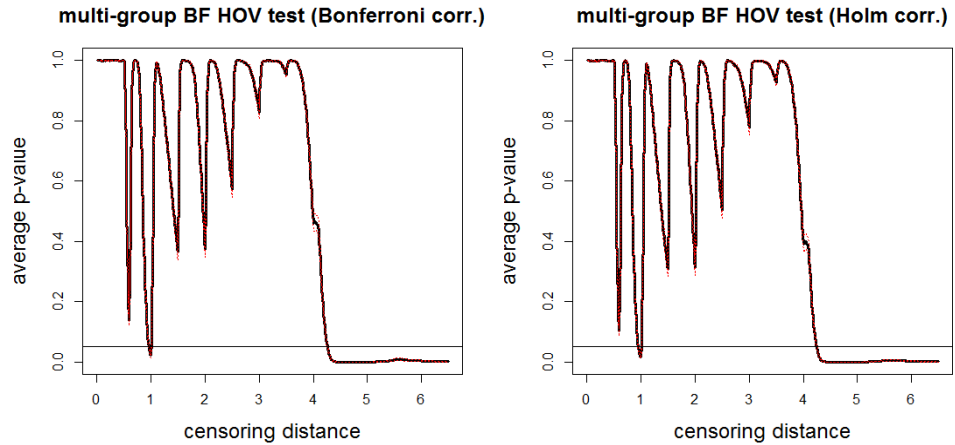

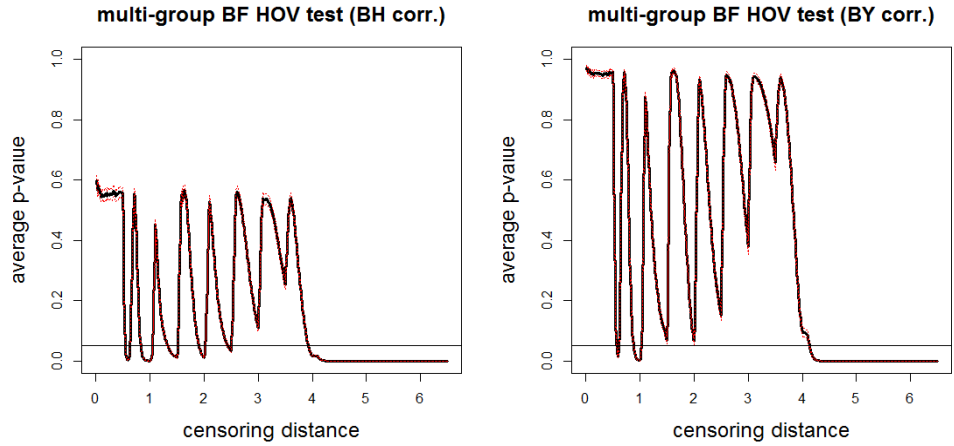

**Figure 10:** Plotted are the average corrected $p$-values (solid line) versus censoring distance values for multi-group BF test together with the 95% confidence bands (dashed lines) based on 10000 Monte Carlo replications of censored $\mathcal{X}$, $\mathcal{Y}$, and $\mathcal{Z}$ sets that are generated under the alternative hypothesis in Equation (6). BH stands for Benjamini-Hochberg correction and BY stands for Benjamini-Yekutieli correction, and "corr." stands for correction.

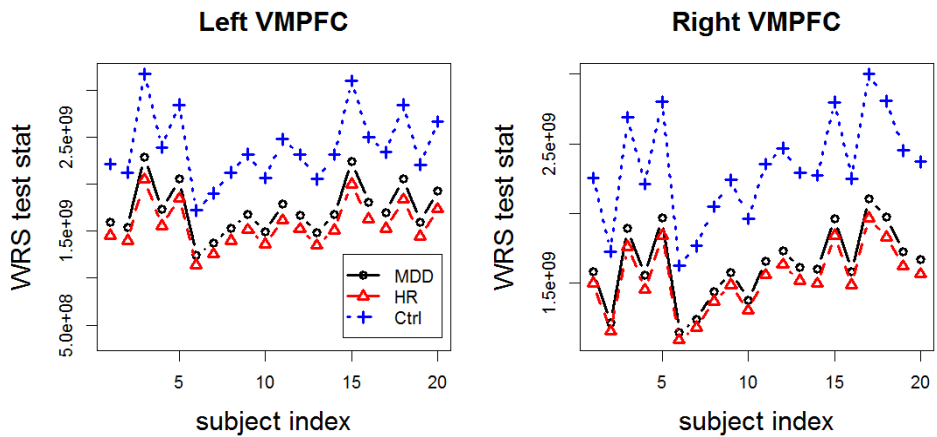

**Figure 11:** WRS test statistic for each subject of HR group compared to HR group (without the particular subject being tested) and MDD and Ctrl groups.

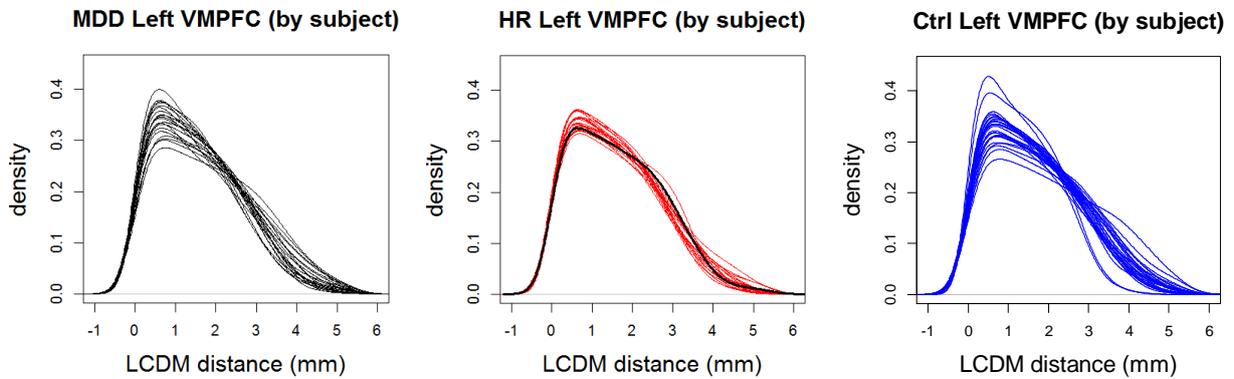

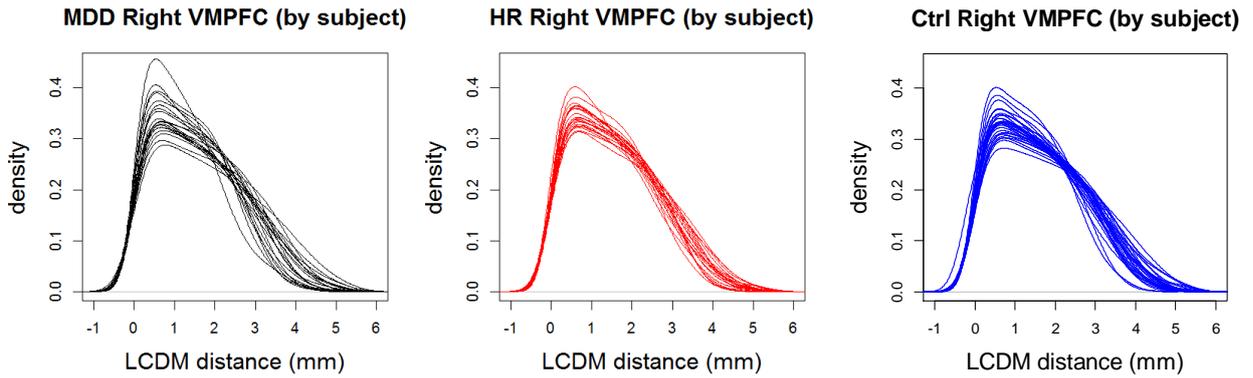

**Figure 12:** Kernel density estimates of LCDM distances for the VMPFCs of the MDD, HR and Ctrl subjects. LCDM distances for each diagnostic group are plotted together on a separate panel. Left VMPFC of HR subject 1 highlighted with a dark solid line.

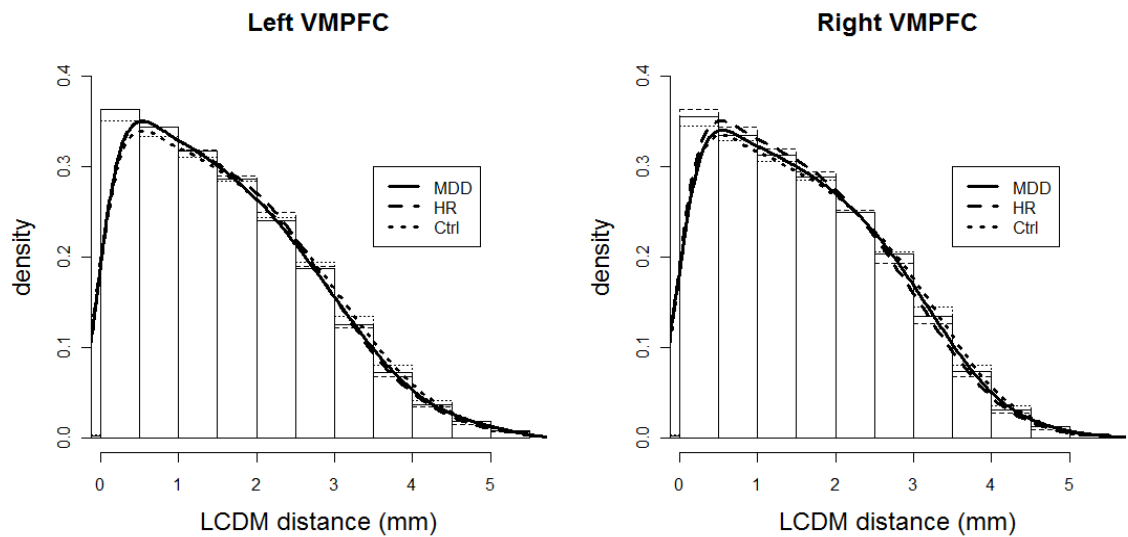

**Figure 13:** Histograms overlaid with the kernel density estimates of the pooled LCDM distances for the left and right VMPFCs.

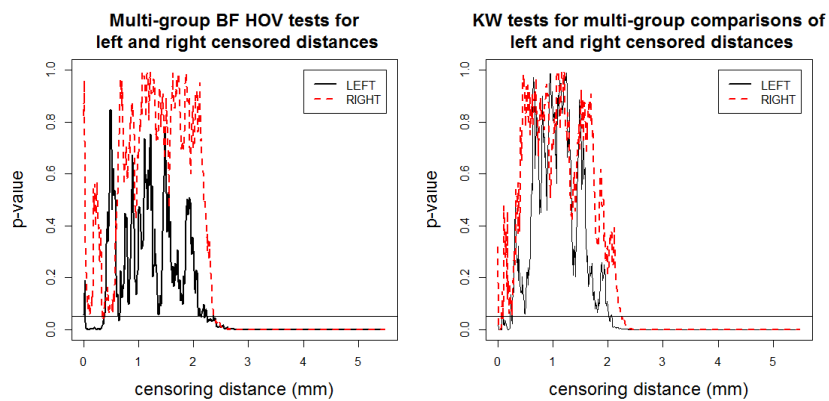

**Figure 14:** The $p$-values versus censoring distance values for multi-group HOV comparison of censored distances of VMPFCs with BF test (left) and for multi-group comparison of censored distances with KW test (right).

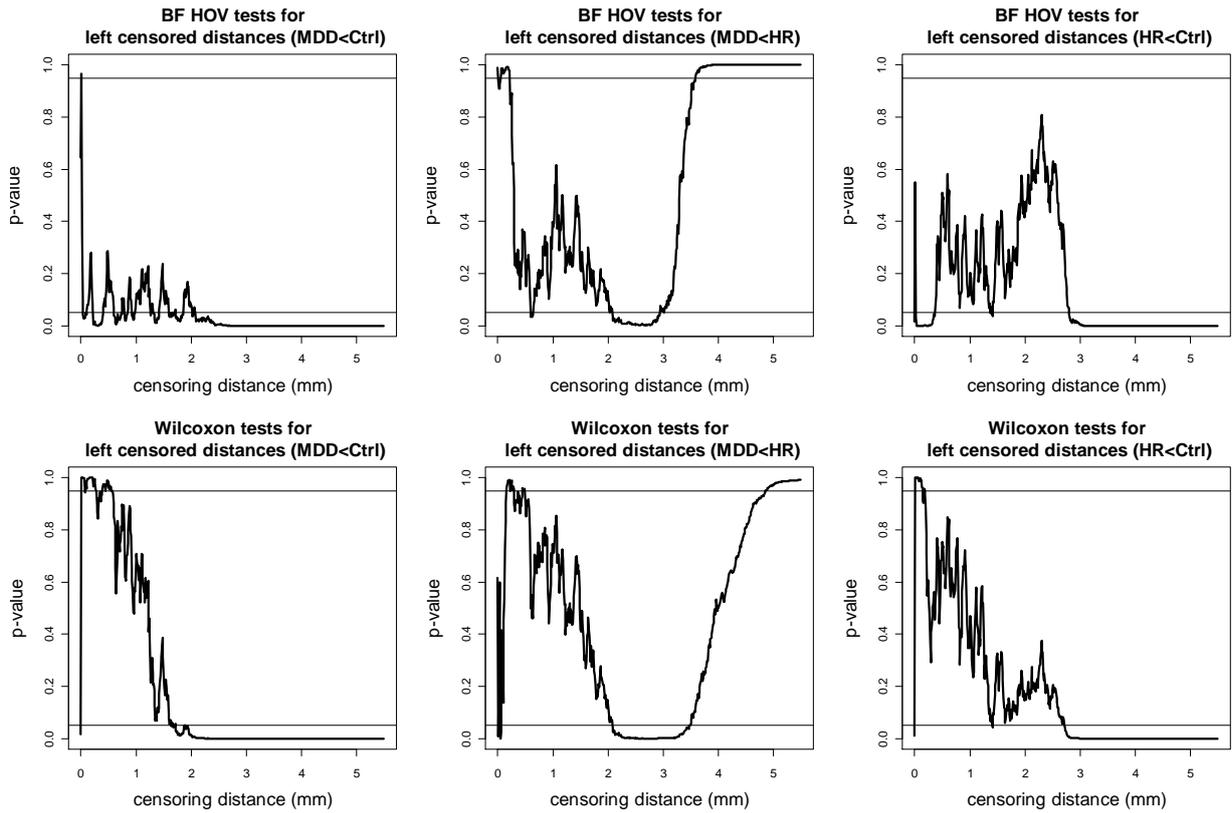

**Figure 15:** The $p$-values versus censoring distance values for pairwise HOV comparisons of left censored distances with BF test (top) and for pairwise comparisons of left VMPFC distances with WRS test (bottom).

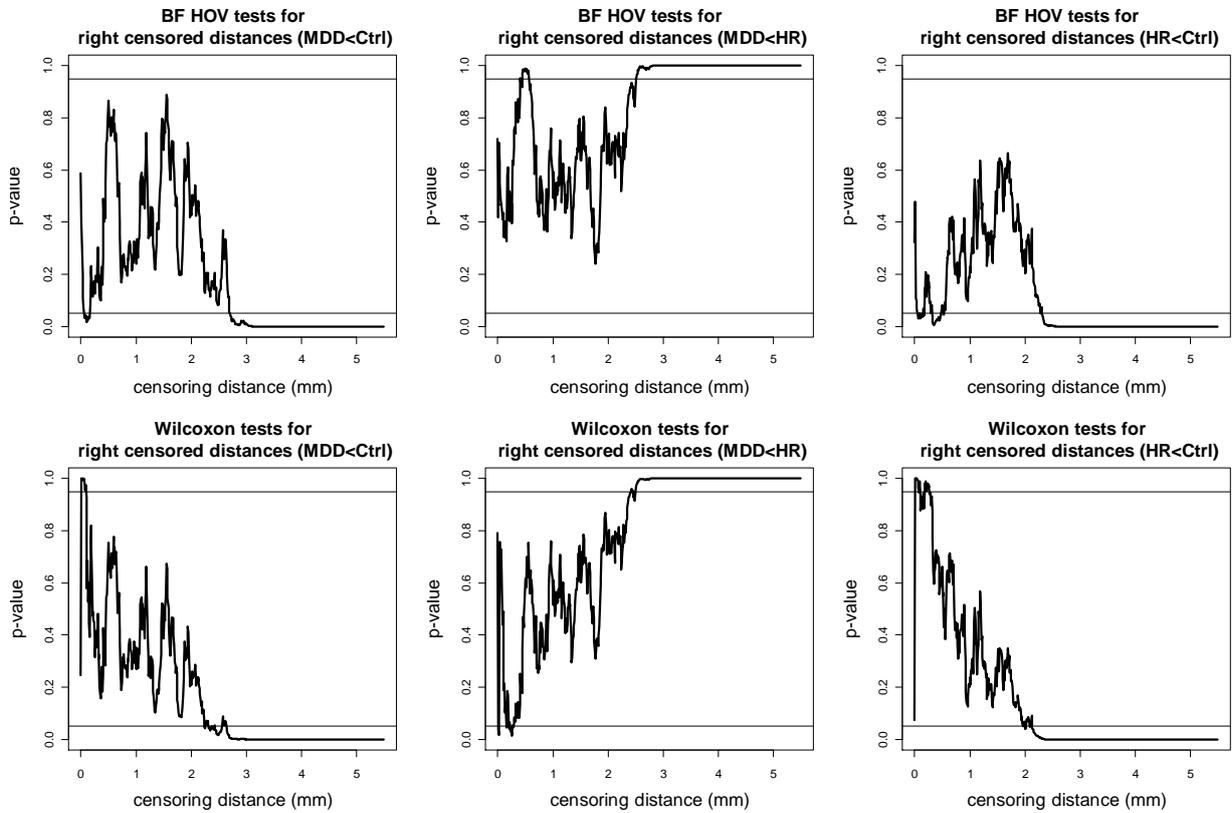

**Figure 16:** The $p$-values versus censoring distance values for pairwise HOV comparisons of right censored distances with BF test (top) and for pairwise comparisons of right VMPFC distances with Wilcoxon test (bottom).